\documentclass{emulateapj}
 
\newcommand{\M}{\mbox{${\cal M}$}}

\newcommand{\msol}{\M$_\odot$}
\newcommand{\lsol}{$L_\odot$}
\newcommand{\hi}{\mbox{H{\sc i}}}

\newcommand{\hh}{\mbox{H$_2$}}

\newcommand{\pa}{$P.A.$}

\newcommand{\kms}{km s$^{-1}$}

\newcommand{\mlb}{$\rm \Upsilon_{Bulge}$}
\newcommand{\mld}{$\rm \Upsilon_{Disc}$}
\newcommand{\vrot}{$v_{\rm rot}$}
\newcommand{\los}{line-of-sight}
\newcommand{\vsys}{$v_{\rm sys}$}
\newcommand{\vobs}{$v_{\rm obs}$}

\slugcomment{Accepted for publication in {\it The Astrophysical Journal}}

\shorttitle{\hi\ kinematics and dynamics of M31}
\shortauthors{Chemin et al.}

\begin{document}

\title{\hi\ kinematics and dynamics of Messier 31\footnotemark[1]}
\footnotetext[1]{Observations obtained at the Dominion Radio Astrophysical
    Observatory (DRAO), operated as a national facility by the National Research Council of Canada}

\author{Laurent Chemin\altaffilmark{2}, Claude Carignan\altaffilmark{3,4},
and Tyler Foster\altaffilmark{5}}

\email{Laurent.Chemin@obspm.fr - Claude.Carignan@UMontreal.CA -
FosterT@BrandonU.CA}

\altaffiltext{2}{G\'EPI, Observatoire de Paris, Section Meudon \& Universit\'e
    Paris 7, 5 Place Jules Janssen, 92195 Meudon, France}
\altaffiltext{3}{Laboratoire d'Astrophysique Exp\'erimentale (LAE),
    Observatoire du mont M\'egantic, and D\'epartement de physique,
    Universit\'e de Montr\'eal, C.P. 6128, Succ. Centre-Ville,
    Montr\'eal, QC, Canada H3C 3J7}
\altaffiltext{4}{Observatoire d'Astrophysique de l'Universit\'e de Ouagadougou (UFR/SEA),
03 BP 7021 Ouagadougou 03, Burkina Faso}    
\altaffiltext{5}{Department of Physics \& Astronomy, Brandon University,
    Brandon, MB, Canada R7A 6A9}

\begin{abstract}

We present a new deep 21-cm survey of the Andromeda galaxy, based on high resolution observations performed with the Synthesis Telescope and 
the 26-m antenna at DRAO. The  \hi\ distribution and kinematics of the disc are analyzed 
and basic dynamical properties are given.
The rotation curve is measured out to 38 kpc, showing a nuclear peak at 340 \kms, a dip at 202 \kms\ around 4 kpc, two distinct  
flat parts at 264 \kms\ and 230 \kms\ 
and an increase to 275 \kms\ in the outermost regions. 
Except for the innermost regions, the axisymmetry of the gas rotation 
is very good. A very strong warp of the \hi\ disc is evidenced.  
The central regions appear less inclined than the average disc inclination of 74\degr, while  
the outer regions appear more inclined.  
Mass distribution models by $\Lambda$CDM Navarro-Frenk-White, Einasto or pseudo-isothermal dark matter halos with baryonic 
components are presented. They fail to reproduce the exact shape of the rotation curve. 
No significant differences are measured between the various shapes of halo. 
The dynamical mass of M31 enclosed within a radius of 38 kpc is 
$(4.7 \pm 0.5) \times 10^{11}$ \msol. The dark matter component is almost 4 times more massive than the baryonic mass 
inside this radius. A total mass of $1.0 \times 10^{12}$ \msol\ is derived inside the virial radius. 
New \hi\ structures are discovered in the datacube, like the
detection of up to five \hi\ components per spectrum, which is very rarely seen in other galaxies.  
The most remarkable new \hi\ structures are thin \hi\ spurs and an external arm in the disc outskirts. 
A relationship between these spurs and outer stellar clumps is evidenced. 
The external arm is 32 kpc long, lies on the far side of the galaxy and has no
obvious  counterpart on the other side of the galaxy. 
Its kinematics clearly differs from the outer adjacent disc.  
Both these \hi\ perturbations could result from tidal interactions with galaxy companions.  

\end{abstract}


\section{Introduction}
In the last decade, several efforts have led to a better understanding of the building history of the disc of  
the Andromeda galaxy (hereafter M31, Table~\ref{opt_par}), the most nearby massive spiral galaxy to the Milky Way (MW). 

One of the most exciting result is probably the discovery of an extended   
stellar halo around M31 \citep{iba01}. More and more studies are completing this new view of the stellar distribution of M31 
and are revealing little by little the complex  chemical, kinematical and morphological nature of its halo 
\citep{fer02, rei02, iba04, iba05,  bro06, guh06, kal06a, kal06b, mar06, far07, iba07, maj07, bro08, cha08}. 
Basically, these authors find that the outskirts of the stellar disc are
 very clumpy and irregular, and the extended halo
is a mix of many metal poor and rich substructures like the south-eastern Giant Stream (GS) and other seemingly shorter streamlike extensions, 
low luminosity dwarf  satellites and globular clusters, as well as an old metal poor underlying (primordial?) halo which extends up to at 
least 150 kpc (in projection). These stellar tidal features and companion galaxies 
 are the probable main imprints  of the hierarchical growth of the M31 stellar disc and halo,   
like the ones encountered in numerical models of dark matter evolution in the framework of 
the Cold Dark Matter (CDM) paradigm \citep[e.g.][]{fre88, kly99, spr05}. 
Numerical models trying to explain the presence of the GS tidal debris around M31 propose a tidal interaction with a $\sim 10^9$ \msol\  galaxy progenitor 
\citep{far06}. 

The gas content in the halo of M31 has also been probed recently from deep \hi\ measurements, allowing the detection  
of a population of discrete low luminosity clouds, with masses in the range
of $10^5$ to $10^7$ \msol\ \citep{thi04,wes05}. The authors suggest that those
clouds could be the analogs of the high-velocity \hi\ clouds seen around the MW \citep{wak99}.
They also suggest a manyfold origin of that  gas (residual from galaxy merger or interaction, cooling flow in the Local
Group, etc...), which may only contribute to $\sim 1$\% of the total \hi\ 
disc mass of M31. 
\hi\ gas clouds from the intergalactic medium could thus also be the building blocks of M31.

At the same time, the advent of the new generation of X-ray, ultraviolet and infrared space observatories allows to probe 
the structure of the interstellar medium of M31 with unprecedented details \citep{wil04,thi05,bar06,gor06,sti08}. 
For instance, the mid-infrared images reveal a dominant ring-like distribution for the  dust component.
Its perturbed morphology seems to have been shaped by the passage  of the
companion Messier 32 through the M31 disc \citep{thi05,blo06}. 
 A high resolution kinematical  survey of the molecular gas contained in the ring and spiral structures has also been presented \citep{nie06}.
 
A large part of the \hi\ studies that have been done on M31 go back up to $\sim$30 years ago 
\citep{gui73, eme74, new77, cra80, unw80a, unw80b, bri84a}. These data were acquired at low spectral and/or 
angular resolution data and did not necessarily cover the whole disc of the galaxy, mainly for sensitivity reasons.
Another more recent \hi\ study of M31 has been presented in \citet{bra90} from VLA observations.
Perhaps the most important result from all these studies is that the \hi\ disc of M31 exhibits a warp whose effects are 
remarkable in the datacube by the presence of many spectral \hi\ peaks along different line-of-sights.  
Another interesting result by \citet{bra91} is the modeling of a rotation curve that  
appears to decline as a function of the galactic distance. 
A study of the mass distribution using this rotation curve only requires stellar bulge plus disc components, 
with no need of any dark matter halo. If it is really the case, M31 would be very different from every other spiral galaxies 
which are known to exhibit a flat rotation curve (or even increasing) 
at large galactocentric distances and thus to contain a massive hidden mass, unless the 
law of Gravity is modified in these acceleration regimes \citep{mil83,mil08}.  

This result was the main motivation to get single dish observations along the photometric major 
axis for the approaching disc half. Indeed, radial velocities of the \hi\ 
gas in the receding side of M31 
are contaminated by \hi\ gas in the MW. 
We concluded that the outer \hi\ rotation curve can not 
be decreasing but seems remarkably flat at the largest radii \citep{car06}. 
However, since it was not possible to model the warp properly with those single dish data, 
it was decided in 2005 to get full 2D velocity information from wide-field synthesis observations at DRAO. 

In this article, we present preliminary morphological, kinematical and dynamical results obtained 
from the analysis of the DRAO \hi\ observations. 
The direct objectives of this article are to present the most extended \hi\ distribution of M31, 
to derive an accurate \hi\ rotation curve in order to verify 
whether the rotation velocities really decrease at large galactocentric distances or really remain constant, as well as
to derive its basic dynamical parameters.

It is worth mentioning here the very recent, wide-field and high angular \hi\ imaging presented in 
\citet{bra09} with the help of the WSRT and GBT telescopes. Future results from that very deep dataset 
 will surely serve as independent comparison with those we present in this article and in other forthcoming 
papers from this series. 

The article is organized as follows. Section \ref{observations} describes
the DRAO observations and the basic data reduction steps. 
Section~\ref{datacube} describes the line fitting procedure 
as well as how the Galactic \hi\ was subtracted from the datacube.
Section~\ref{hicontent} describes the \hi\ content and distribution
while section~\ref{sed:hikinematics}  concentrates on the
kinematics and on the calculation of the \hi\ rotation curve. A comparison 
with results from different studies  is also done in this section. 
An analysis of the perturbed outer regions of the disc is done in section~\ref{discussion-hioutskirts}.
 Finally, a study of mass distribution models is done in section~\ref{massmodels}. 
Concluding remarks are given in section~\ref{conclusion}.

All velocities are given in the heliocentric rest frame and a 
Hubble's constant of H$_0$ = 73 \kms\ Mpc$^{-1}$ is chosen throughout the article \citep{spe07}. 
Since the symbol $R$ refers to as the galactocentric radius and most of the magnitudes and luminosities   are given in the  photometric 
$R-$band, no subscripts are attached to magnitudes or luminosities
 for clarity reasons, except where explicitely mentioned.

\begin{table}[!t]
\caption{Basic optical parameters of M31.\label{opt_par}}
\begin{tabular}{ll}
\tableline\tableline
Parameter & Value \\
\tableline
Right ascension (J2000)$^a$ &  $00^{\rm h}42^{\rm m}44.4^{\rm s}$ \\
Declination (J2000)$^a$ &  $+41\degr16\arcmin08\arcsec$ \\
Morphological type$^a$ & SA(s)b \\
Distance$^b$ (kpc) & $785 \pm 25$ \\
& (1$\arcmin$ = 229 pc) \\
Systemic Velocity$^a$ (\kms)& $-300$ $\pm 4$   \\
Optical radius$^a$, $R_{25}$ & 95.3$\arcmin$   \\
Inclination$^c$ & 77$\degr$   \\
Position angle$^c$ & 35$\degr$    \\
Total apparent $B$ magnitude$^c$ & 4.38   \\
Corrected total $B$ magnitude$^d$ & 3.66  \\
Absolute $B$ magnitude & $-20.81$ \\
Total blue luminosity &
$3.1 \times 10^{10}$ \lsol   \\
(\bv)$^c$ & 0.91   \\
(\ub)$^c$ & 0.37   \\
($B-R$)$^e$ & 1.37   \\ 
\tableline
\end{tabular}
\tablenotetext{a}{\citet{dev91}} \tablenotetext{b}{\citet{mcc05}}
\tablenotetext{c}{\citet{wal87}} \tablenotetext{d}{$A_g$ = 0.33 ,
A(i) = 0.8 log (R$_{25}$) = 0.39} \tablenotetext{c}{\citet{wal88}}
\end{table}

\section{Observations and Data Reduction}
\label{observations}

\subsection{HI emission line observations}
A total of five fields towards M31 were observed in the 1420~MHz continuum and
21~cm line with the Synthesis Telescope (ST) at the Dominion Radio
Astrophysical Observatory (DRAO) between September and December 2005.
This interferometer consists of seven
9~metres diameter antennae along an East-West baseline 617.1~m long. The primary beam
of each element is 107$\arcmin$.2 (FWHM), and structures down to the resolution
limit of $58\arcsec \times 58\arcsec/\sin(\delta)$ in the 1420~MHz line
are resolved within this beamwidth (a Gaussian taper to the $u,v$ data is
applied, broadening somewhat the synthesized beamwidth up from the nominal
$49\arcsec \times 49\arcsec/\sin(\delta)$ resolution in the continuum).
Further instrumental details on the DRAO ST are found in \citet{land00}.

Tables~\ref{tab:obsparam} and \ref{tab:centres} list the observational parameters and 
field-of-view centres for the five individual aperture synthesis observations. 
Fields were chosen such that a spacing of $\Delta=$77$\arcmin$ between centres was observed, giving
nearly equal sensitivity over the whole area surveyed. The exposure
times per field was 144 hours.

 \begin{table}[!t]
\caption{Observational parameters of the DRAO aperture synthesis observations of Messier 31.\label{tab:obsparam}}
\begin{tabular}{ll}
\tableline\tableline
Parameter & Value \\
\tableline
Observation dates & 2005 September-December  \\
Total length of observation &  144$\times$5 hrs \\
Velocity centre of band & -300 \kms \\
Total bandwidth & 4 MHz (843 \kms) \\
Number of velocity channels & 256 \\
Frequency sampling & 15.6 kHz   \\
Velocity resolution & 5.3 \kms   \\
Number of spatial pixels$^a$  & 1024\\
Pixel angular size$^a$ & 22\arcsec \\
\tableline
\end{tabular}
\tablenotetext{a}{Values given for the full  resolution initial datacube.}
\end{table}
\begin{table}[!t]
\caption{DRAO Synthesis Telescope observational set-up and noise. \label{tab:centres}}
\begin{tabular}{c|c|c|c}
\tableline\tableline
Obs. & Field centre & Beam parameters &  Noise at field \\
date & coordinates & $\theta_{maj}(')\times\theta_{min}(')$   &  centre\\
(2005) & (J2000, h:m:s,$\degr$:':") & and orientation ($\degr$) &  (K)\\
\tableline
Oct.~05 &
0:52:19.7,~43:15:00&
1.46$\times$0.98,~$-$90.2&
1.12\\
Sep.~09&
0:47:32.0,~42:19:00&
1.50$\times$0.98,~$-$90.2&
0.98\\
Sep.~09&
0:42:44.3,~41:16:09&
1.53$\times$0.98,~$-$90.0&
0.83\\
Nov.~04&
0:37:56.6,~40:08:00&
1.58$\times$0.97,~$-$89.8&
1.04\\
Dec.~10&
0:33:18.9,~39:00:00&
1.57$\times$0.97,~$-$87.2&
1.06\\
\tableline
\end{tabular}
\medskip

Summary of 1420~MHz line + continuum observations centered on and
surrounding Messier 31. Noise is indicated for the full resolution initial datacube.
\end{table}

\hi\ emission line images are made in each of 256 channels across a
4~MHz bandwidth centered on $v =-$300 \kms. The spectrometer gives a resolution of 5.27 \kms, and
each channel samples a width of $\Delta v=$ 3.3 \kms. The
\hi\ line datacube spans 843 \kms. The theoretical noise at the
pointing centre is 1.75 $\sin(\delta)$~K, corresponding closely to
the measured values in Table~\ref{tab:centres} (for channels free from \hi\ emission).

To depict structures accurately in the radio continuum and \hi\
line images (especially those of angular size $\geq$ 56$\arcmin$, missed by the
interferometer due to its shortest baseline limit of 12.9~m), low spatial
frequency information is routinely added to each ST map. These
``short-spacing" data are obtained from \hi\ line observations made
with the DRAO 26-m paraboloid \citep{hig00}. The flux is corrected for stray
radiation entering through the sidelobes, and the continuum is subtracted from
all channels before the integration with the ST observations. The spatial
resolution of these data is 37$\arcmin \times$35\farcm3; all spectrometer
settings were the same as for the ST data.

\subsection{Data Reduction and Mosaicing}
\label{sec:datareduction} Prior to mosaicing our five fields
together, we perform some standard data processing steps  
\citep[processing methods are similar to those developed for the
Canadian Galactic Plane Survey CGPS, described in][]{tayl03}.
First, a continuum baseline level is removed from each datacube.
Average emission line-free channels at both the low and high
velocity ends of the cube are made, and a linear interpolation
(determined from these two continuum maps)  is then subtracted from
each channel map in the datacube. To calibrate the flux scale of
each cube, we compare point sources in the average of the two
continuum end-channels to those in the 30~MHz continuum band map of
each field (these maps are first CLEANed around the strongest 
sources). This can be done since continuum maps are flux calibrated
against several strong sources that are routinely observed by the ST
\citep[e.g. 3C~48, 3C~286; see Table 1 of] [for sources and
fluxes]{tayl03}. An error-weighted mean of the flux ratio $S_{\rm
1420~cont}/S_{\rm HI~cont}$ for all point-sources above a cutoff
level (20~mJy~beam$^{-1}$) is obtained; this sample is further
trimmed by rejecting sources $\pm$0.5 away from the mean. Typically,
20-30 sources remained in the sample for each field. The uncertainty
in the flux calibrated this way is $\sim$5\%.

The processed fields are combined into a 1024 square pixels
mosaic, centered on optical coordinates
$RA=\rm 00^{h}42^{m}44.3^{s},DEC = +41\degr~16\arcmin~09\arcsec$
(J2000). The pixel size is 21\farcs875. The central 93$\arcmin$
radius of the primary beam constitutes the final width of the five
individual fields, which were mosaiced together. The final mosaic
spans 515$\arcmin$ oriented along M31's disc. The final noise values
for the mosaic (measured) are $\Delta T_{b}\sim$0.85~K in the
individual field centres, and $\Delta T_{b}\sim$0.95~K in the
overlapping regions.

The final step consists in merging interferometer data with single-dish
(26-m telescope) observations. For simplicity, this
merging is performed in the map plane \citep[as opposed to the $u,v$ visibilities
plane, as is done for the CGPS; see][]{tayl03} using the procedure
described below. The interferometer mosaic is first convolved to the
resolution of the single-dish mosaic; this is then subtracted from
the 26-m mosaic. The resulting difference map should contain \hi\
structures invisible to the interferometer, so it is added back to
the full-resolution ST mosaic. This procedure gives equal weight to
low and high spatial frequency structures, and while it is somewhat
different from the method used in the CGPS (where a tapering
function weights the overlapping spatial frequencies of both maps),
the results of both approaches are very similar for \hi\ line
images.

A spatial  binning of 2$\times$2 pixels is finally applied to the datacube 
in order to decrease noise in the spectra.
The datacube has a final dimension of 512$\times$512 pixels, with a pixel size of 43\farcs 75.
This angular sampling is well sufficient for the kinematical and dynamical purposes. It corresponds to 167 pc
at the distance of 785 kpc. The DRAO synthesized beam size of $\sim 60\arcsec \times 90\arcsec$ samples
a linear scale of $\sim 230\times 340$ pc in the \hi\ disc of Messier 31. It is
as resolved as a typical \hi\ mapping of 10 Mpc distant discs observed
with a high resolution beam resolution of e.g. 5\arcsec. 
The present \hi\ observations can thus be considered as high resolution ones and 
for this reason no correction for beam-smearing is needed.  
 
\begin{figure*}
\begin{center}
\includegraphics[width=\textwidth]{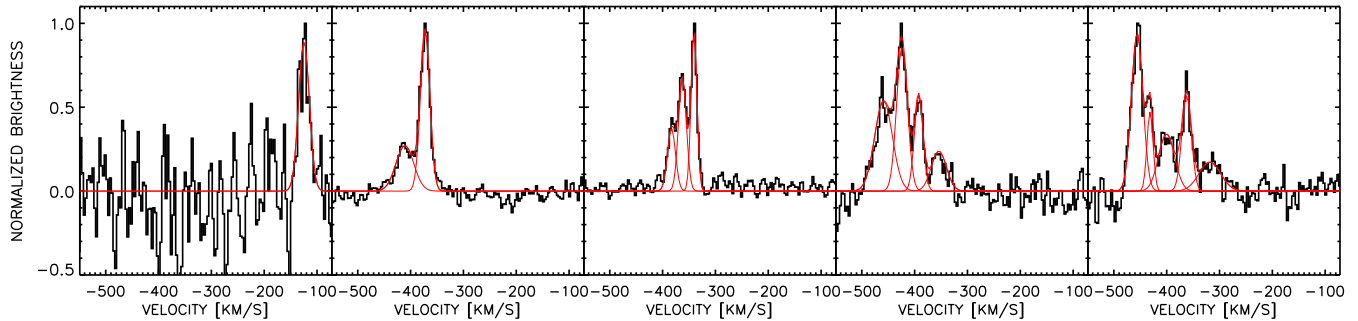}
\caption{Selected M31 spectra from the \hi\ datacube showing multiple \hi\ peaks  
 and fitted components (red solid lines).}
\label{fig:m31spectrafit}
\end{center}
\end{figure*}

\section{\hi\ datacube analysis}
\label{datacube}

\subsection{Previous works}

 \hi\ spectra of Messier 31 are known to
exhibit multiple peaks in emission \citep[e.g.][]{new77,cra80,baj82}. 
It is also the case in molecular gas observations \citep{dam93,loi99,nie06}.
For the \hi\ data, this is particulary well illustrated along position-velocity (PV)
diagrams made parallel to the major axis in \citet{cra80} or 
\citet{bri84b}, where a second velocity peak is seen in addition to
the main (brighter) peak. Both these lines draw a tilted
figure-eight shape in which the main peak traces the usual rotating
pattern of the disc (steep velocity rise in the inner regions sometimes 
accompanied by a flatter velocity part in the external regions) and the second peak
shows a linear (shallower) velocity rise all along the slices.

With the noticeable exceptions of \citet{bri84a}, \citet{bri84b} 
or \citet{bra91}, the previous \hi\ studies of M31 analyzed their data with 
a single emission line approximation 
in order to derive an integrated emission map, a velocity field
or a rotation curve. 
For instance, \citet{new77} reported that when two peaks are detected, 
the radial velocity of the spectrum is chosen at the barycentre of the lines.
Though it seems a reasonable hypothesis in regard to the low spectral 
resolution of all these old observations, it surely provides a
biased velocity distribution for M31. Integrated
fluxes and velocities are uncorrectly estimated when 
lines are blended.

A simple explanation for the origin of two components comes from the
fact that the \hi\ disc of Messier 31 is warped at large
galactocentric radius. Because the disc is highly inclined, the projection effects enable us to cross two times the
disc along the \los\ (\hi\ gas is supposed to be optically
thin). The \los\ velocities are thus composed of a main \hi\
component lying in the internal disc region in addition to a secondary
component lying in the external warped region of the disc but 
seen projected at small radii.  The presence of the
warp were the reasons that led 
\citet{bri84b} to analyze the WSRT data differently from the
other studies by a warped and flaring \hi\ layer model. 
They proposed two velocity fields, one  for the
disc and another one for a separate, external warped structure. From this, they
deduced that about 39\% of the total \hi\ mass of the disc could
reside beyond 18 kpc, in its warped region.

\subsection{Current analysis}
\label{sec:fitdata}

\subsubsection{Evidence for multiple HI spectral components}

A provisional study of the DRAO datacube has been presented in
\citet{car07} where moments maps were derived by fitting a single
gaussian to each spectrum of the datacube. We extend here this simplistic view of our data 
to a more detailed analysis where multiple \hi\ components are now fitted to the spectra. 
The high-sensitivity observations indeed reveal a new more complex view of the neutral gas in M31 than what has  
been shown in  past \hi\ studies.
Figure~\ref{fig:m31spectrafit} displays several \hi\ spectra picked at
different positions through the datacube. The \hi\ spectra exhibit
almost all the time more than one component, sometimes up to 5 peaks. 
This result confirms the basic view shown in e.g. \citet{bri84b} where two \hi\ components are shown (and 
sometimes more), but largely extends the detection of new multiple spectral components in the datacube. 
One clearly wants to emphasize here that a single or two peak analysis cannot apply straightforwardly 
to the current data. This would provide biased intensity, velocity and velocity dispersion maps.

\subsubsection{Contamination by the Milky Way}
\label{sec:mwcontamination}

\begin{figure}[!b]
\includegraphics[width=\columnwidth]{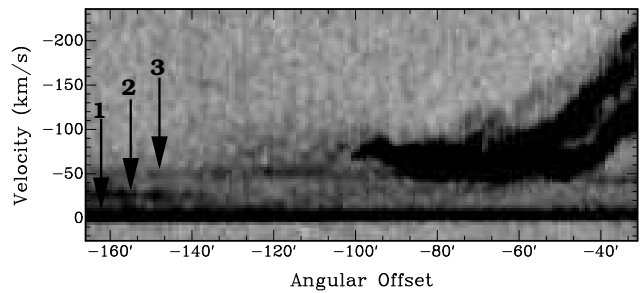} 
\caption{Contamination of the data by Galactic \hi. The slice through the datacube shows 3 \hi\ components from
the Milky Way at $\sim -4$ \kms\ (1), $\sim -20$ \kms\ (2) and $\sim
-40$ \kms\ (3). }
\label{fig:mwcontamination}
\end{figure}

A usual problem with observations of M31 is the contamination from the MW. 
M31 is so massive that its receding half can reach radial velocities that coincide with 
those of the Galaxy.  However most of Galactic \hi\ remains relatively easy to detect and to model  
  because its lines can be observed almost 
everywhere in the field of view (Fig.~\ref{fig:mwcontamination}). We detect two major Galactic lines. A
first bright component is found at $-4 \pm 2$ \kms\ for a
velocity dispersion of $6 \pm 3$ \kms\ as deduced from 66000
spectra.  This line does not contaminate the emission from Messier 31 due to its
too high radial velocity. Another line is found at $-40 \pm 11$
\kms\ for a velocity dispersion of $18 \pm 10$ \kms, as deduced
from 19500 spectra free from M31 \hi\ emission. 
This is the line that mostly contaminates the NE
 half of M31.
Another minor Milky Way line is observed around 
$-20$ \kms\ (see Fig.~\ref{fig:mwcontamination}). 
This emission line does not contaminate the
signal from M31 because of its high radial velocity. Other very faint lines 
only revealed by very high contrast imaging are found around $-70$ \kms\ and $-90$ \kms. Their 
contribution to the whole emission is very negligible.

The method used to subtract the MW emission from the datacube 
is similar to the one employed by \citet{bra09}. 
For each channel map, a mask containing pixels with \hi\ emission 
of M31 is first created. The mask covers a small fraction of the field-of-view 
and allows to blank the \hi\ emission of M31.
 A two-dimensional model of the Galactic emission is then 
generated by fitting a 3$^{\rm rd}$ order 2D-polynomial to the blanked channel map. 
The 2D model is then subtracted from the initial channel map.
This method allows to remove most of the galactic \hi. 
Residual MW contamination is removed by hand during the cleaning of the 
multiple velocity fields (see \S\ref{sec:emlinesorting}). 
We tried another approach to remove MW \hi\ by simultaneously 
 fitting gaussian lines in addition to \hi\ from M31. However   
 that subtraction method had to be rejected because 
 too many negative residuals are created in the cube, introducing 
 artificial local minima in the integrated emission map.
 
As claimed in \citet{bra09} it is very difficult to define whether  
small scale \hi\  structures observed around the disc of M31 are bound  
 to Andromeda or simply Galactic cloudlets. For simplicity reasons, we decide 
 to ignore the gas emission out of the main disc in the following analysis.

There is no clear mention in older \hi\ studies of M31 
 of  Galactic emission around $\sim -40$ \kms.   It is thus possible that 
measurements of the \hi\ flux and velocity dispersions are overstimated 
for the receding half of the disc in those works. However their velocity fields 
remain very marginally contaminated.

\begin{figure}[!t]
\includegraphics[width=\columnwidth]{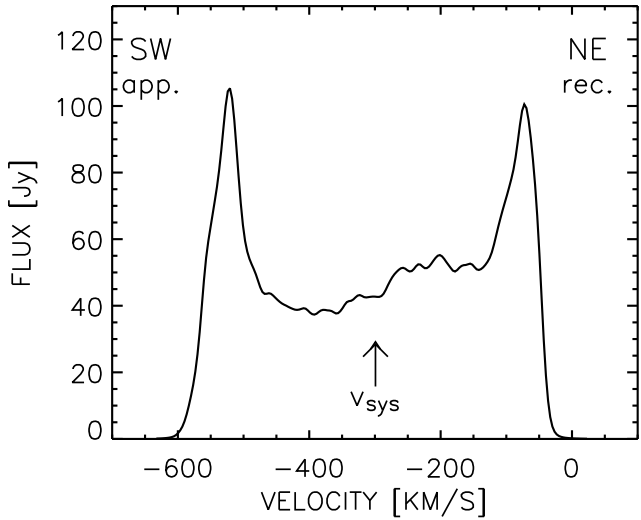} 
\caption{\hi\ integrated profile of Messier 31. \vsys\ refers here to as the
integrated weighted mean systemic velocity derived from the profile.}
\label{fig:higlobal}
\end{figure}

\subsubsection{Emission line detection and fitting}
\label{sec:emlinesorting}

The detection of emission lines is done by keeping all channels above a 3$\sigma$ level, where $\sigma$ is 
the intensity dispersion derived from half  
of the spectrum. This dispersion measurement 
is justified by the fact
that most of the channels are free from \hi\ emission. The method
to locate a peak in the remaining channels is a usual slope
computation technique. An emission line is detected when the slope sign does not change
 along at least two consecutive channels, as measured both to the 
left and right sides of a peak. Once detected, velocities and amplitudes of
these peaks serve as initial guesses for the minimization step where 
gaussian profiles are fitted to the spectrum. A ``continuum"
value is fitted as an additional parameter as well. It is very close to zero because a baseline
  has been subtracted in each pixel during the data
reduction  (see \S\ref{sec:datareduction}). Boundary constraints are applied
to the parameters (amplitude, velocity centroid and dispersion). 
A  velocity centroid
has to be fitted within the observed spectral range, a velocity
dispersion has to be greater than one channel width and lower 
than an artificial (unexpected) large value of 130 \kms\ and 
a peak amplitude has to be greater than zero
 and lower than the highest intensity observed in a spectrum.
 Results of spectral fittings of 1 to 5 
 \hi\ components are displayed in Fig.~\ref{fig:m31spectrafit}.

A final cleaning step is then applied to the data. All fitted lines whose
 amplitude is larger than a threshold of 3$\sigma$ were
kept.  All lines whose velocity dispersion is greater than 60 \kms\ are
discarded because such high values turn out to be very rare and, above all, unrealistic.
Obvious residuals of MW emission are also removed from the different maps.
As a result of this filtering procedure, among the 25350 available spectra in the \hi\ datacube 
and in which at least one component is detected with a signal-to-noise ratio of at least 3, two peaks are 
 detected in $\sim$66\% of them, three peaks in $\sim$35\% of them,  four peaks in
 $\sim$14\% of them  and five peaks in  $\sim$3\% of them. A large number of 
 \hi\ components ($>$ 3 peaks) is therefore rare in the datacube.

\begin{figure}[!b]
\includegraphics[width=\columnwidth]{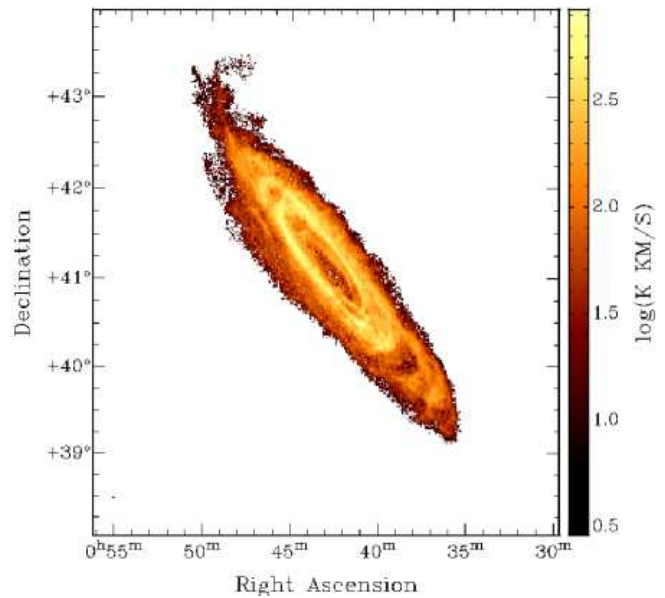} 
\caption{Total \hi\ distribution of Messier 31. The small ellipse in the bottom-left corner displays the angular  size of the synthesized beam 
($\sim 1\arcmin \times 1.5\arcmin$).}
\label{fig:hitot}
\end{figure}

\section{HI distribution and kinematics}
\label{hicontent}

\subsection{Integrated properties}

\begin{figure}[!t]
\includegraphics[width=\columnwidth]{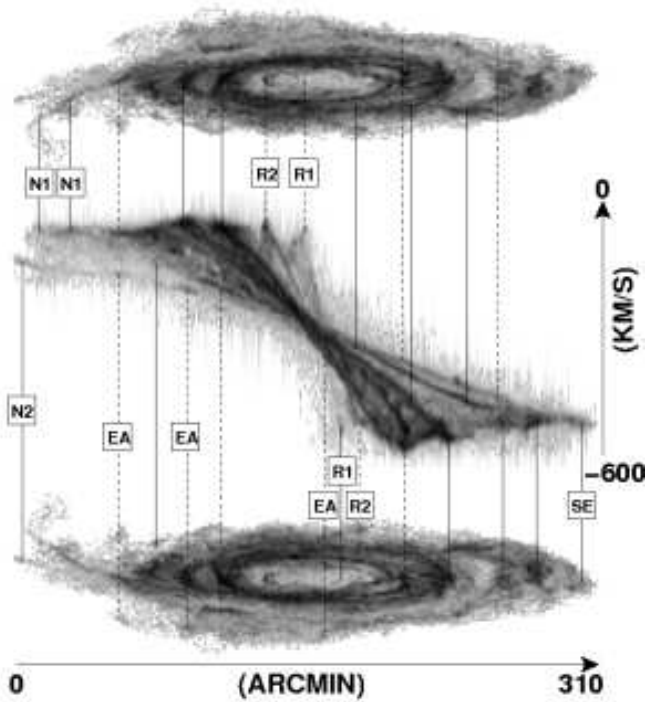}
\caption{3D view of the \hi\ datacube of M31.  
The middle panel is the position-velocity diagram of the \textit{whole} datacube projected 
on the photometric major axis, which is aligned along the horizontal axis. 
The bottom panel is the total integrated \hi\ image. The  top panel is the mirror image of 
the bottom panel. Solid (dashed) lines are for \hi\ emission from the front (respectively far) side of the galaxy. 
Labels N1, N2, SE show the two northern \hi\ spurs and the south-western extension, EA the external arm, R1 and R2 the 
ring-like structures in the centre of the disc (see text for details, \S\ref{sec:3D}).}
\label{fig:3dview}
\end{figure}

The flux in each individual channel was summed to give the global
\hi\ profile of  Figure~\ref{fig:higlobal}. The profile is relatively regular. 
 The two peaks are almost perfectly symmetric. 
An asymmetry is nonetheless observed between them. There is more gas in the receding side 
of the galaxy than in its approaching part. An intensity-weighted
systemic velocity of $-299 \pm 3$ \kms\ and a midpoint
heliocentric radial velocity of $v = -297 \pm 3$ \kms\
are derived.  The measured profile widths
at 20\% and 50\% levels are $W_{20} = 533\pm 3$ \kms\ and
$W_{50} = 509 \pm 3$ \kms. This can be compared to the RC3
\citep{dev91} values of $v = -300 \pm 4$ \kms, $W_{20} = 536
\pm 7$ \kms\ and $W_{50} = 510 \pm 7$ \kms, which show a good 
agreement with ours within the errors.

The integrated flux of $29221.4$ Jy \kms\ implies a total \hi\
mass of $4.23 \times 10^9$ \msol. 
This mass is larger by 15\% than the one measured in \citet{bri84b}, 
is comparable with the estimate by \citet{cra80} and
 is smaller by 25\% than the one derived in \citet{bra09} (before opacity corrections). 
The difference between all mass estimates is likely 
due to the various sensitivities of the observations and to the accuracy 
of the subtraction of the contamining MW emission.

\subsection{The 3D  structure of M31}
\label{sec:3D}
 
Figure~\ref{fig:hitot} displays the total  \hi\ integrated emission of M31 and  
a 3D view of the datacube is shown in Fig.~\ref{fig:3dview}. 
The lowest \hi\ column density of the DRAO observations at a 3$\sigma$ 
peak detection level is $\sim 1.7 \times 10^{19}$ cm$^{-2}$  
and the highest \hi\ column density is $\sim 5 \times 10^{21}$ cm$^{-2}$. 
 The maximum disc extent is $R \sim$2.6\degr\ ($R = 36.4$ kpc), as derived along the photometric major axis.
It corresponds to 1.67 $R_{25}$ and 6.48 $R_d$, with 
 $R_d$ being the stellar disc $R-$band scale-length derived in \S\ref{sec:stelpotential}.

The high resolution \hi\ map reveals a disc with very little gas in its central regions, 
a feature usually observed in early-type discs.
Two ring-like structures are observed around  $R \sim 10\arcmin$ (2.5 kpc) and $R \sim 20\arcmin-25\arcmin$ (4.6-5.7 kpc). 
We refer to them as R1 and R2 in Fig.~\ref{fig:3dview}. At such high inclination it is difficult to firmly claim whether they are real rings,  
like e.g. those created by gas accumulation at the location of inner or ultra-harmonic Lindblad resonances,  or 
tightly wound spiral arms. They are coincident with dusty ring-like structures observed in NIR images 
from \textit{Spitzer/IRAC} data \citep{bar06} as well as with molecular gas ring-like structures \citep{nie06}. 
Another wider bright ring-like structure is observed for $40\arcmin < R < 80\arcmin$ (9.1-18.3 kpc).  
Though it is often referred to as the ``10-kpc ring" of M31, its morphology is more complex than a regular ring
because holes are observed in it, as seen e.g. in the approaching side of the disc.  Long spiral arms are then clearly seen at large radii.  
 The most prominent spiral arm of M31 is observed in the south-western half of the disc. 

A faint external spiral arm (label EA of Fig.~\ref{fig:3dview}) is discovered on the edge of the receding half of the disc (southern part of the 
NE quadrant of Fig.~\ref{fig:hitot}). It is connected to the long spiral arm which arises from the SW and its apparent end 
is clumpy (see region around $\alpha_{2000} = \rm 00^h 49^m 41.80^s, \delta_{2000} = +42^d 12\arcmin 54\arcsec$). 
Other new structures that were not seen in old \hi\ data are the two disc extremities to the SW and NE (labels ``SE" for south-western extension 
and ``N1"/``N2" for northern spurs in Fig.~\ref{fig:3dview}).
Here thin and faint gaseous extensions are observed.  
Their gas distribution and kinematics are discussed in more details in \S\ref{discussion-hioutskirts}.

The top and bottom panels of Fig.~\ref{fig:3dview} is the right ascension-declination view of the datacube 
while the middle panel is a rotation of this later by 90\degr\ with respect to the  horizontal axis (disc photometric major axis). 
It is thus a position velocity plot of the whole datacube projected onto the major axis. 
Vertical lines are drawn to guide the eye in order to link the kinematics to the spatial distribution of the \hi\ gas. The main features of the diagram are :
\begin{itemize}
\item 
All spiral- and ring-like structures do not cross at the same location in the centre of the diagram ($v = -300$ \kms). 
This is probably caused by a lopsided nature of the disc.
\item
Except for the external arm (EA), all spiral- and ring-like structures have a symmetric counterpart with respect to the galaxy centre. 
This symmetry is not regular in velocity amplitude within $\sim$ 50\arcmin\ around the centre, 
as seen for instance with the velocity peaks of the ring-like structure R2. 
This is also caused by the disc lopsidedness, as well as other probable noncircular gas motions in the central parts.
\item 
A steep velocity gradient is observed in the innermost ring-like structure (R1).
\item
The northern spur N2 appears as a kinematical extension of the external arm in the velocity space.
 
\end{itemize}

The neutral gas distribution of the DRAO observations is consistent with \hi\ images from many other works
\citep{cra80,rob75,baj82,eme74,bri84a}.  The agreement is better with the deep WSRT observations of \citet{bra09} 
because of the comparable sensitivity with the DRAO observations.
Their \hi\ image do show the external arm as well as the north-eastern spurs and south-western extension. 
It thus leaves no doubt about their presence in the disc outskirts of M31. 

\begin{figure*}[!t]
\includegraphics[width=0.2\textwidth]{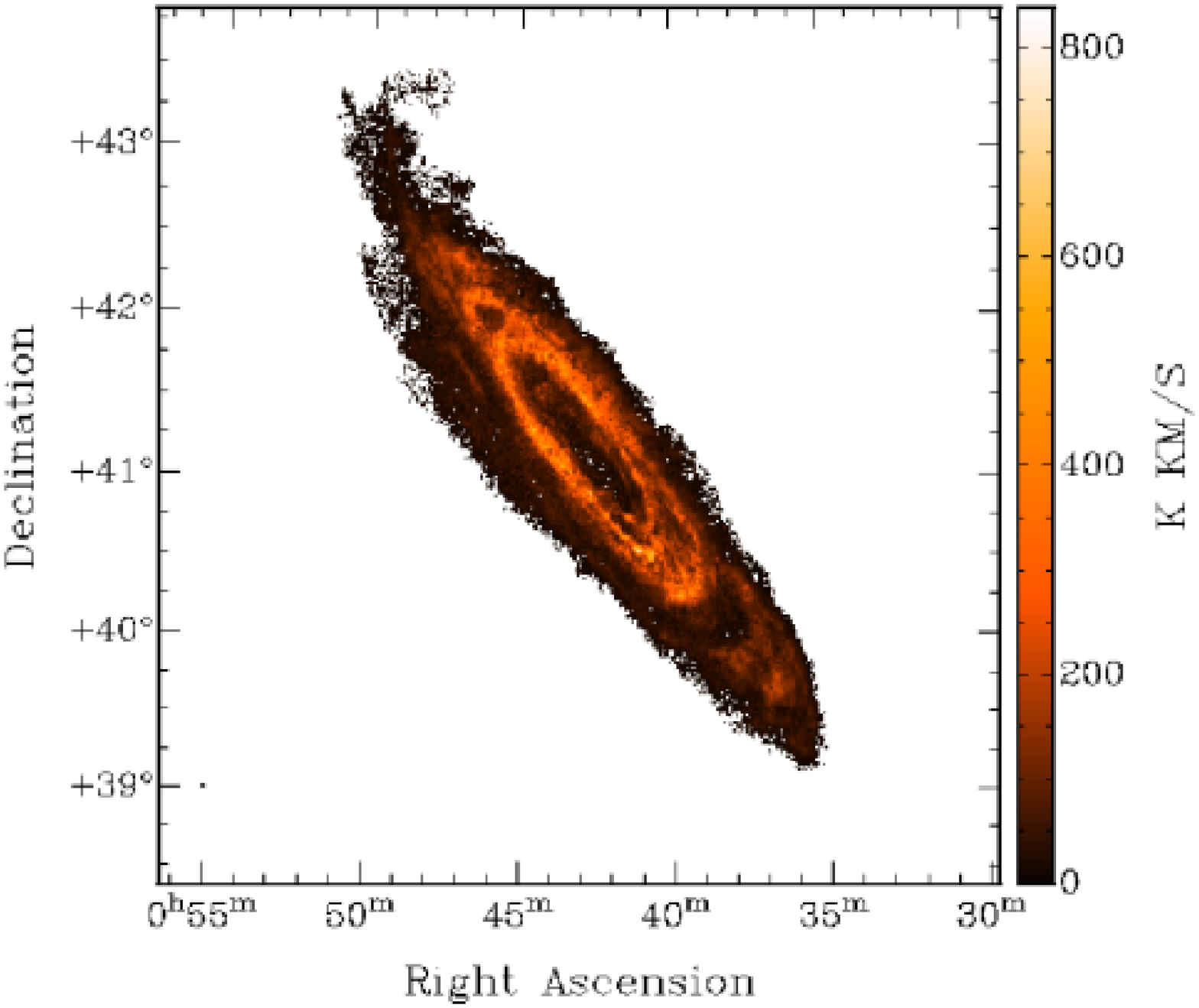}\includegraphics[width=0.2\textwidth]{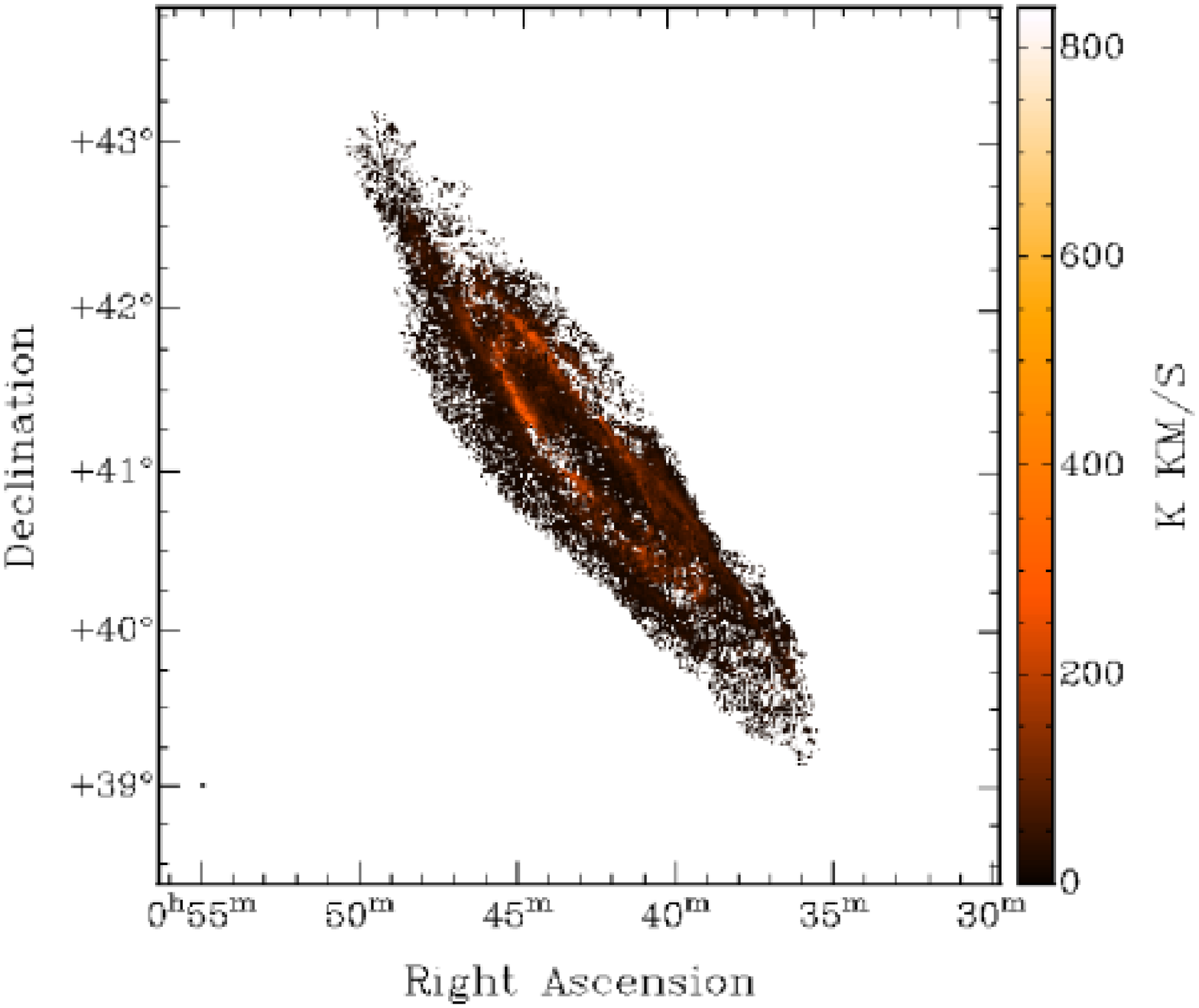}\includegraphics[width=0.2\textwidth]{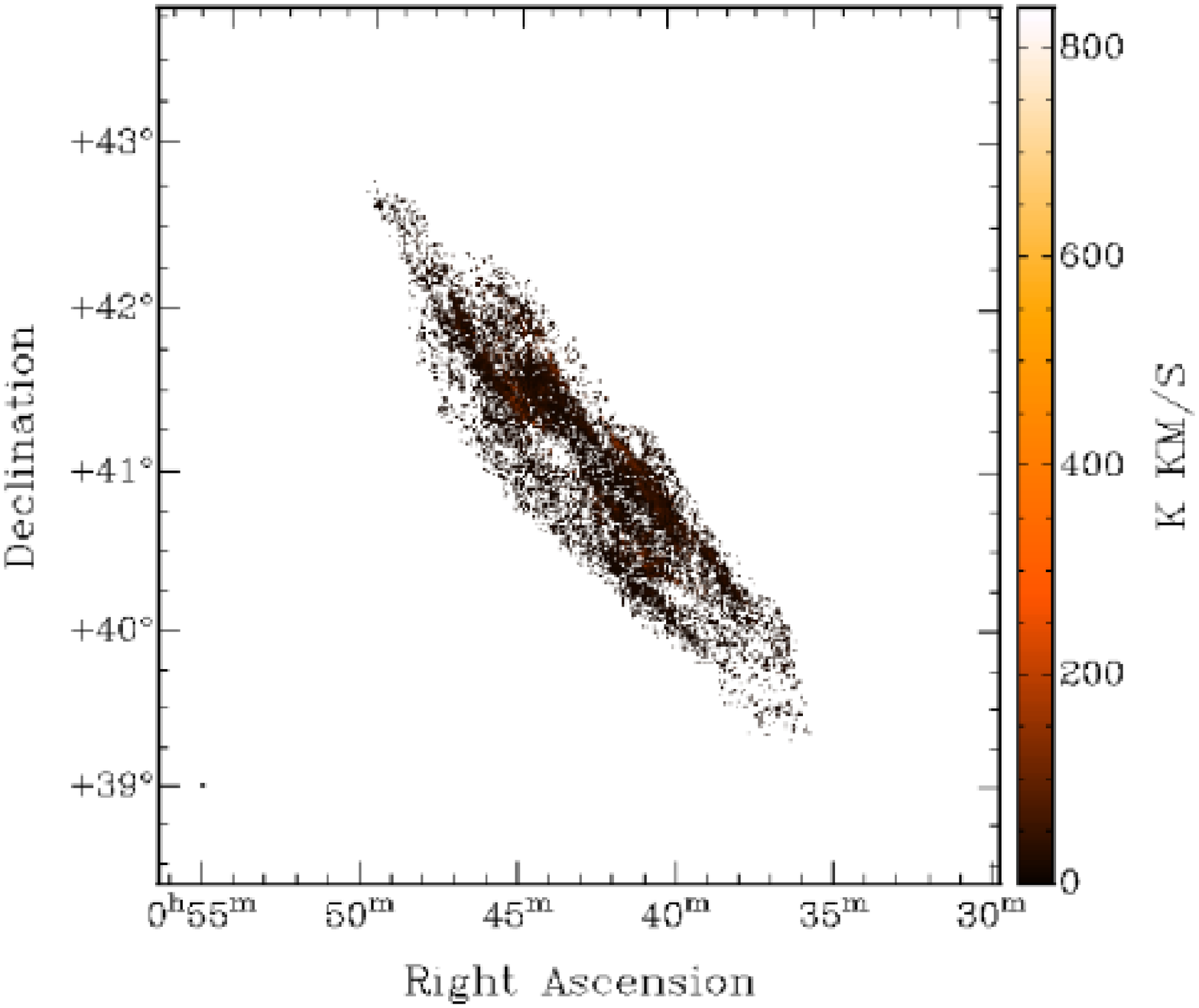}\includegraphics[width=0.2\textwidth]{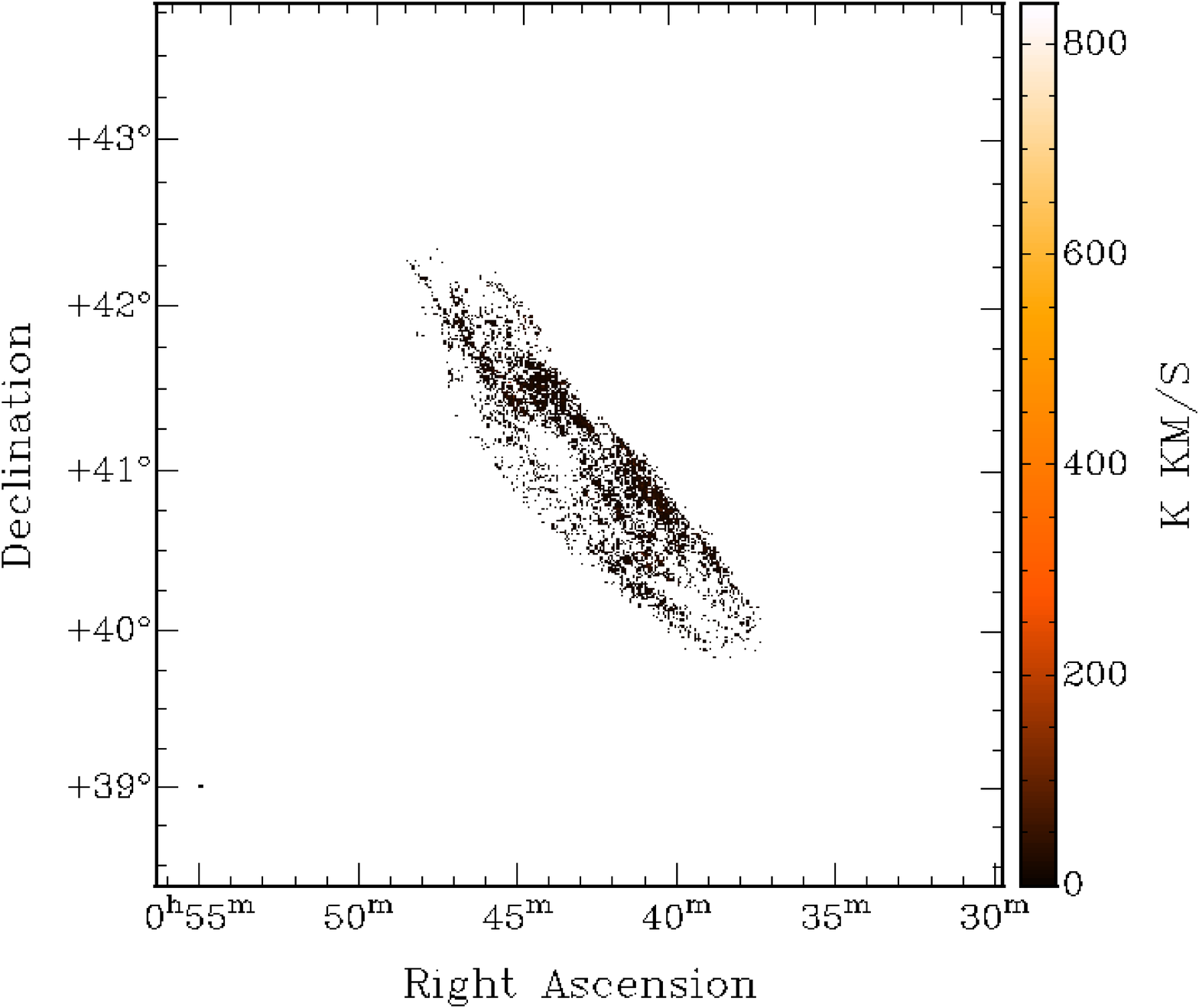}\includegraphics[width=0.2\textwidth]{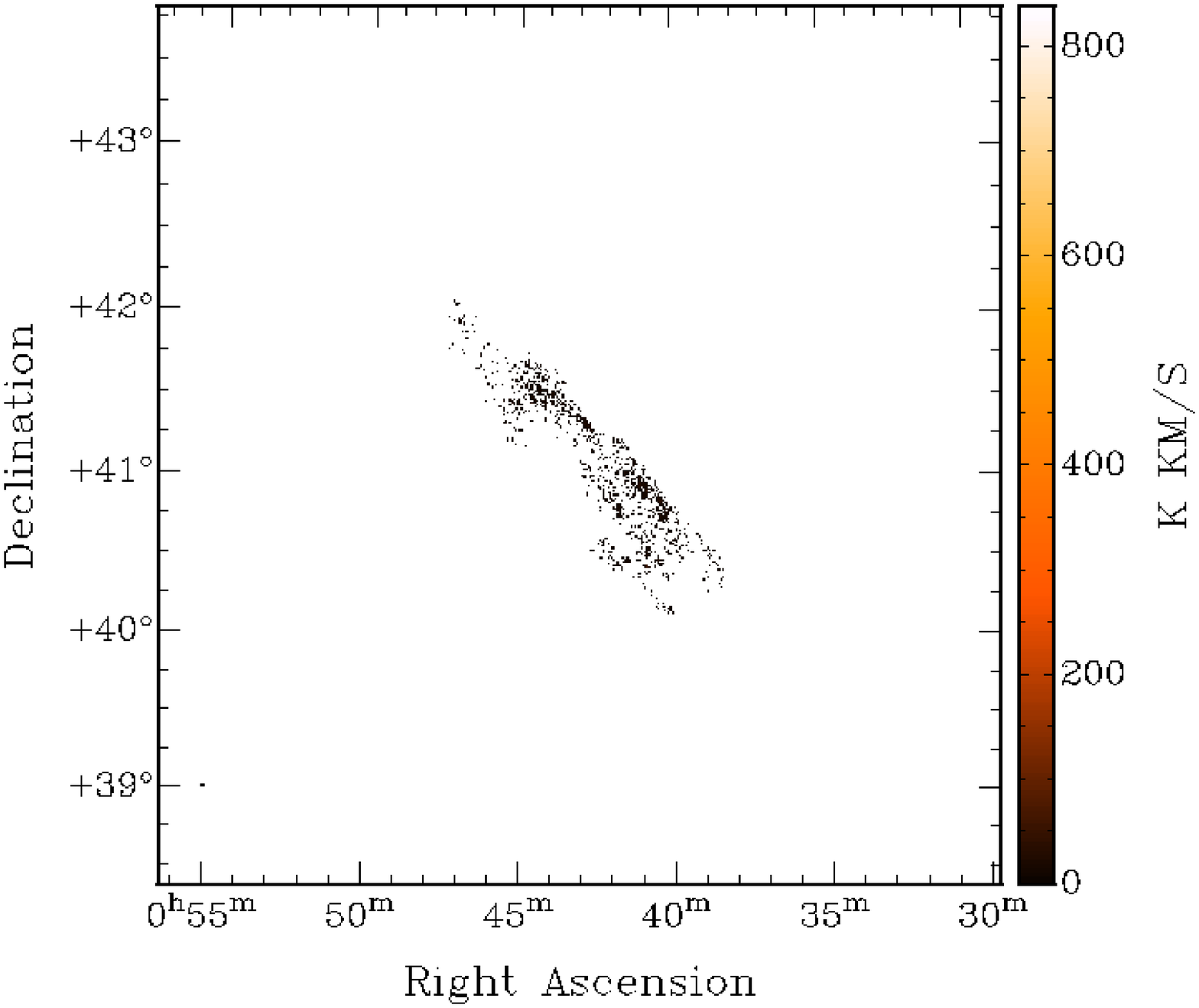}
\includegraphics[width=0.2\textwidth]{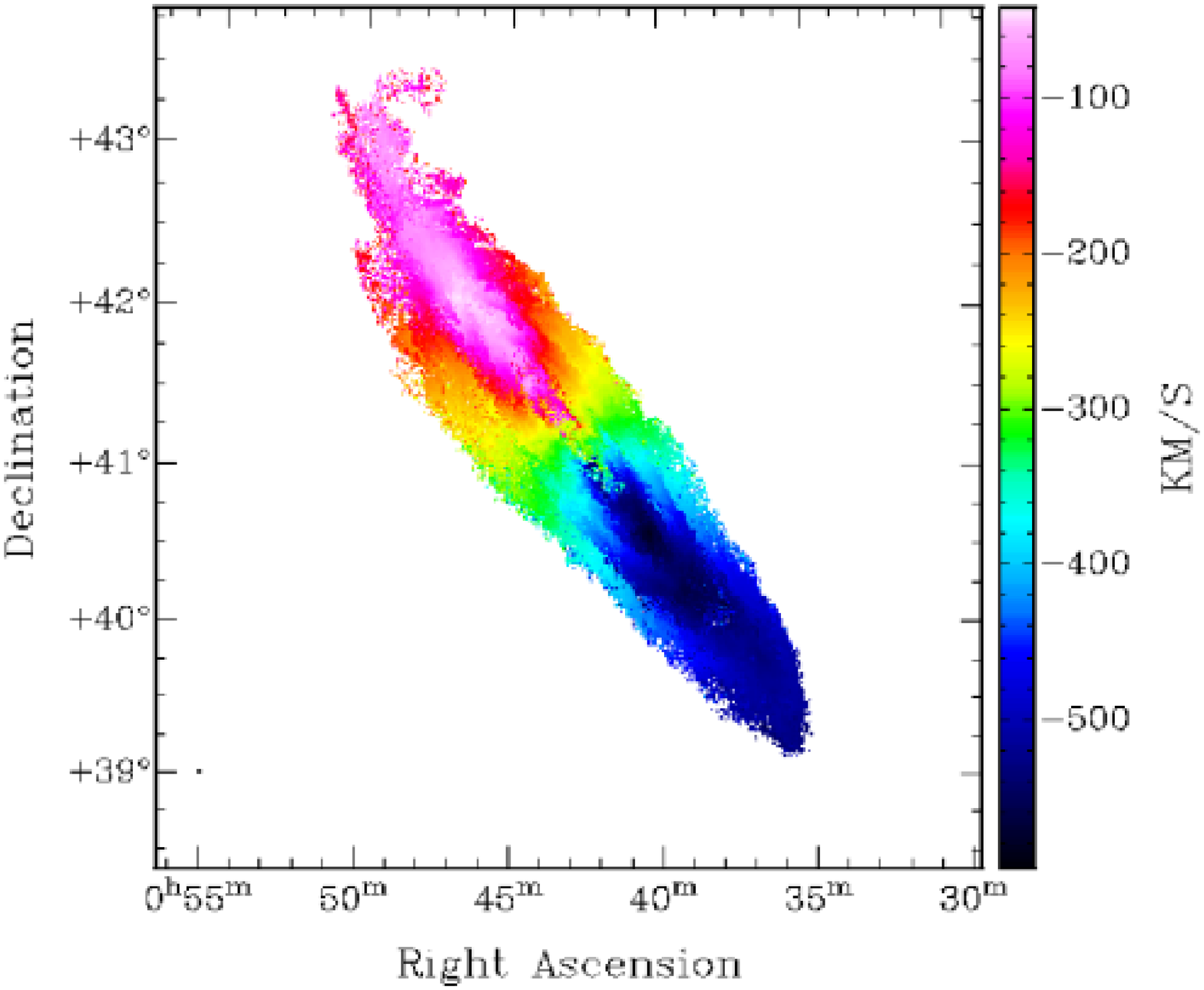}\includegraphics[width=0.2\textwidth]{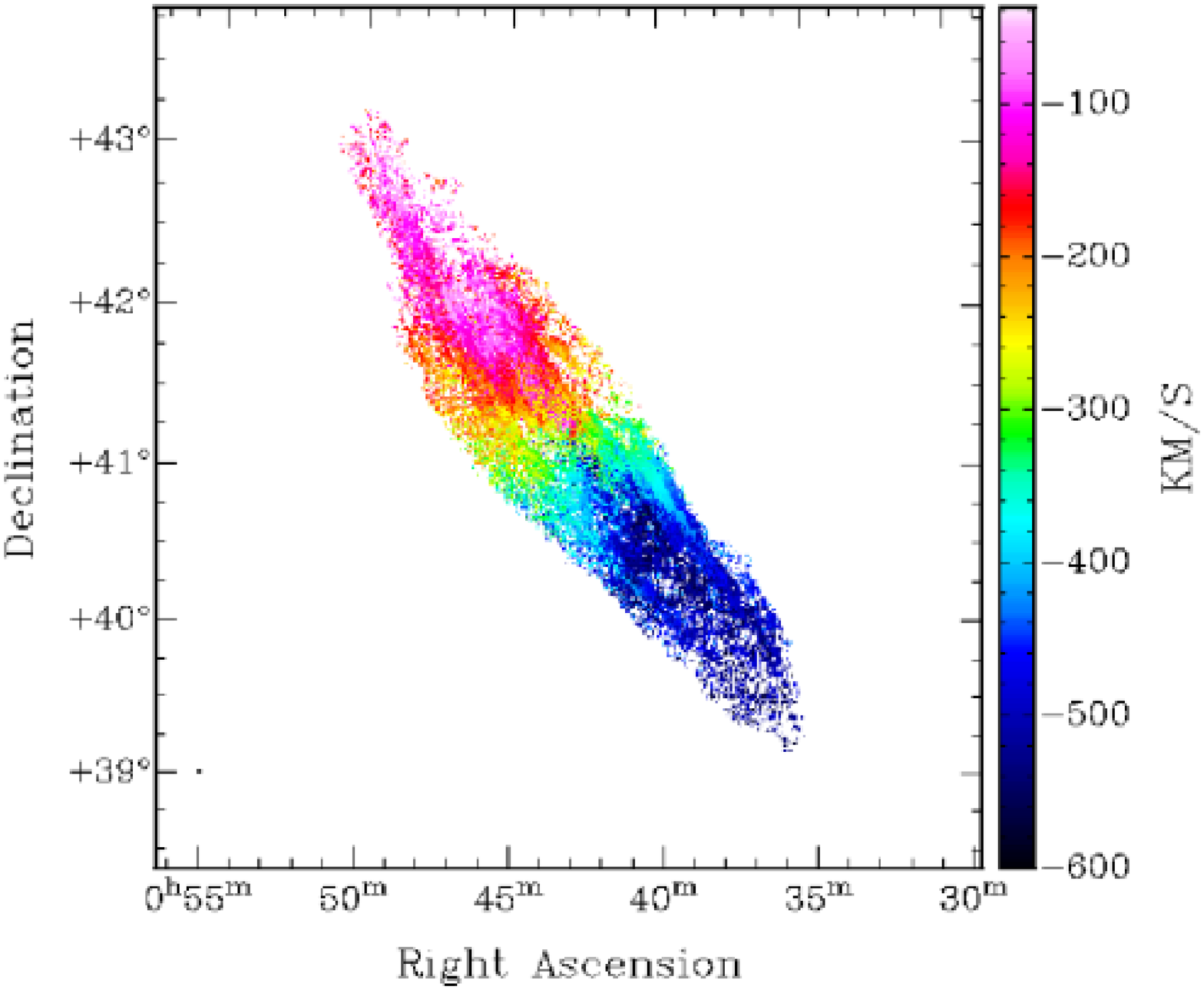}\includegraphics[width=0.2\textwidth]{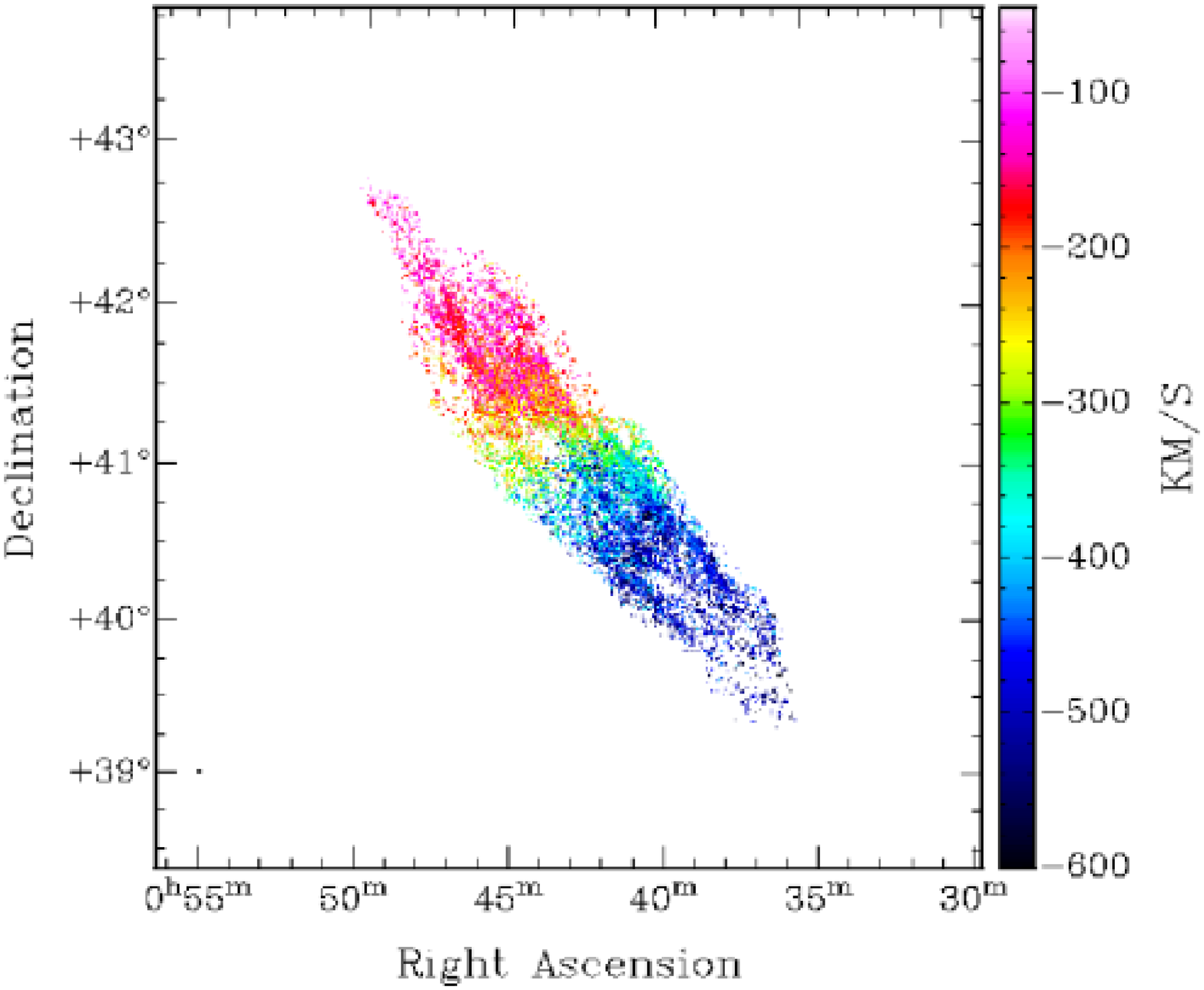}\includegraphics[width=0.2\textwidth]{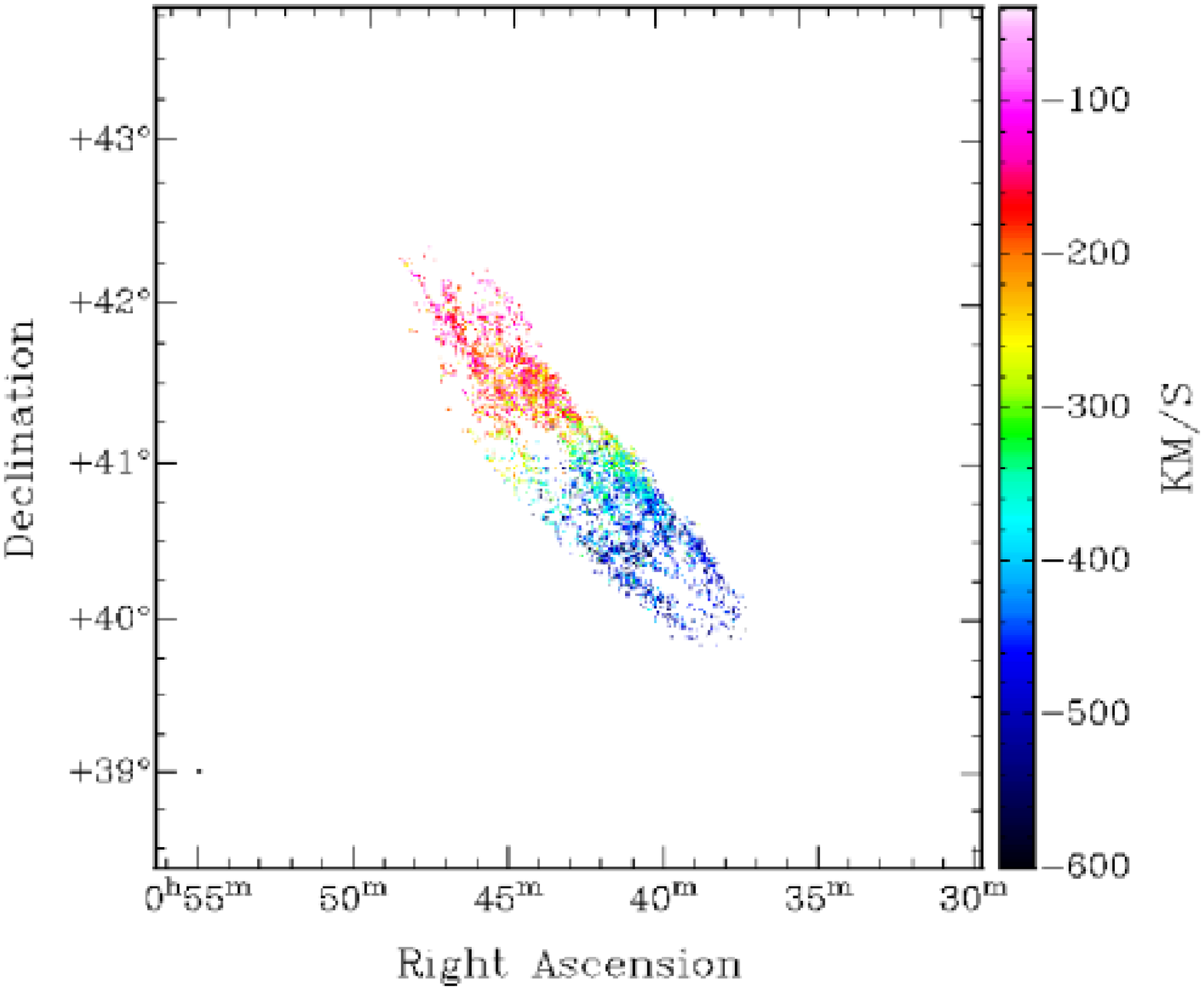}\includegraphics[width=0.2\textwidth]{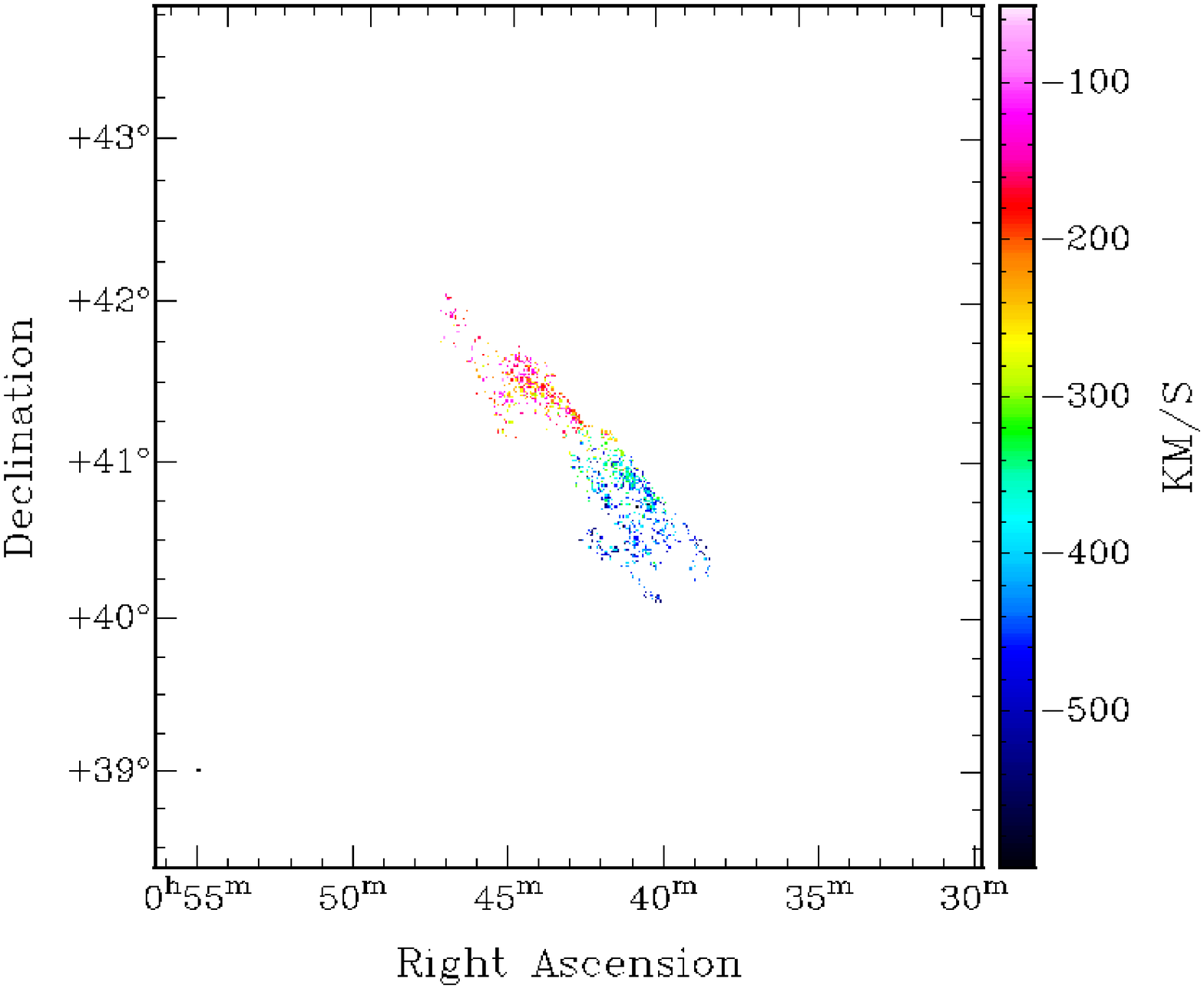}
\caption{\hi\ integrated emission and velocity maps of M31. The \hi\ components are sorted by decreasing brightness from left to right.}
\label{fig:hivf1}
\end{figure*}

The gas distribution for all detected \hi\ components is shown in Figure~\ref{fig:hivf1}. In each pixel, the components are  
displayed by decreasing integrated emission from left to right. A brief discussion on the origin(s) of all 
these \hi\ peaks is proposed in \S\ref{sec:otheremissionline}

\section{Kinematical analysis}
\label{sed:hikinematics}

In \citet{bri84b}, a 3D model of  WSRT observations was presented in order 
to reproduce the complex structure of the datacube. They 
proposed a flared and warped disc model for M31 that can  explain 
the presence of double \hi\ peaks observed in most of pixels. 
Not all of the dozen free parameters of their model are directly fitted to 
the datacube. Some assumptions deduced from the observations had to be made. For instance, 
the modeled rotation curve they use and that comes from the bulk velocities observed along the photometric
major axis is flat all through the inner disc (except in the central regions), 
with a maximum velocity of $\sim 250$ \kms. 
Another 3D model derived from the same WSRT data and \hi\ observations of \citet{eme74}  
is presented in \citet{bra91}. It describes the gas and velocity structures in terms 
of spiral density waves and shows a central disc which is tilted by $\sim$15$\degr$ from 
the median plane of the galaxy.  Modeled position velocity diagrams reveal
 a complex velocity structure \citep[see Fig. 5 of][]{bra91}. 

A complete 3D analysis of the DRAO datacube taking into account both a warped and flaring disc, with spiral 
and/or other density waves, in addition to other processes like e.g. a lagging halo, as observed in other galaxies \citep{fra01,fra02,bar05,oos07}, 
is beyond the scope of this article that aims at presenting preliminary dynamical results from more simple geometrical and kinematical
hypothesis. We only analyze the kinematics of the bulk rotation of M31 by fitting a tilted-ring model to a velocity field, similarly 
to what is usually done in many other  extragalactic studies. The  inclination of the disc  is 
very well suited for such a model  because the degeneracy between the rotation velocity and 
the inclination is small around 75\degr\ \citep{beg89}.

\begin{figure}[!b]
\includegraphics[width=\columnwidth]{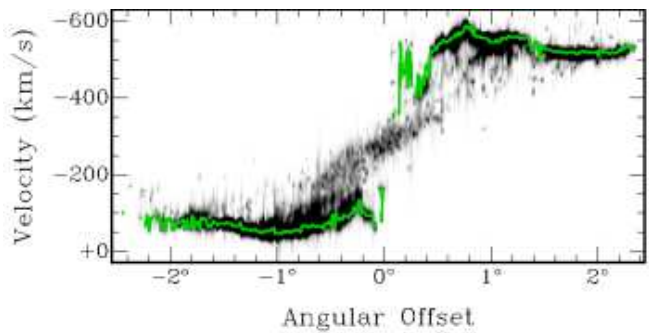} 
\caption{Position velocity plot made along the photometric major axis $P.A. = 38\degr$. A green curve represents a slice 
through the velocity field of the main \hi\ component.}
\label{fig:pvd}
\end{figure}

\begin{figure}[!t]
\includegraphics[width=\columnwidth]{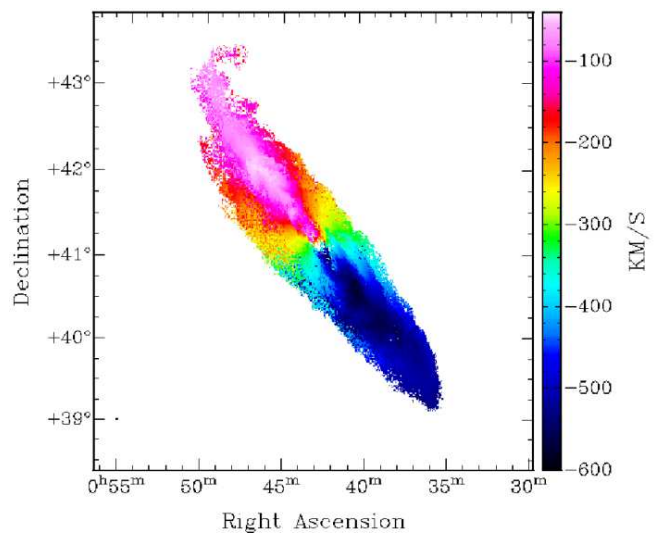} 
\caption{Velocity field of M31 used for the titled-ring model analysis.}
\label{fig:vftitled}
\end{figure}

Probably the major ambiguity with our analysis comes from that it seems difficult to decide for
 pixels that exhibit more than one peak which of the spectral components is the best tracer 
of the bulk disc rotation.  Several methods can be applied to sort the different lines and build a useful velocity map.
For instance, lines can be sorted by their amplitudes, integrated fluxes, velocity centroids or widths. 
There is generally one line whose peak amplitude or integrated intensity strongly dominates the spectrum. 
However choosing the brightest  peak for the bulk rotation such as what is shown in Fig.~\ref{fig:hivf1} 
remains problematic in pixels where the \hi\ emission is dominated by other structures than the bulk disc (like extraplanar gas) 
or when the projection effects contaminates the emission. In particular, the \hi\ emission in the  central regions 
is dominated by gas from the warped part of the disc. 
This is illustrated in the position velocity plot made along the major axis of the galaxy \citep[Fig.~\ref{fig:pvd}; see also][]{bri84b}. 
The steep innermost gradient of the faint central ring-like structure is dominated by the shallow linear emission from the warped gas.
Choosing this brightest later line would lead to  discontinuities in the velocity field and
thus to an erroneous inner rotation curve. 
For each pixel a good compromise is to select the component that has the largest velocity relative 
to the systemic velocity of the galaxy while avoiding isolated faint features.  
It permits rejection of the bright component from that shallow linear emission, which is closer to the systemic velocity, 
and the components beyond the flat part of the position-velocity diagram, which come from possible extraplanar gas (e.g. high velocity clouds). 
The velocity field that is generated following that method is shown in Fig.~\ref{fig:vftitled} and a velocity cut along its major axis 
is reported in Fig.~\ref{fig:pvd} as a green line. 
That map probably allows to obtain the best velocity continuity 
with no 3D model of the datacube. Hereafter we refer to the ``main component" the \hi\ line that has been chosen to derive this velocity field. The  mass 
attached to the main component is $2.49 \times 10^9$ \msol.

\subsection{Tilted-ring model}
\label{sec:tiltedring}

The kinematical parameters and rotation curve are determined by fitting a tilted-ring model to
the velocity field following the procedure described in e.g. 
\citet{ver01} or \citet{che06}. Briefly, the
\textit{rotcur} task \citep{beg89} of the GIPSY  data
reduction software \citep{vdh92} is used to fit the variation of kinematical
parameters as a function of radius. It is considered here that gas
motions are along circular orbits. Hence, no axisymmetric
radial motions (gas inflow or outflow) or vertical motions
perpendicular to the galaxy plane are derived so that sky plane
velocity \vobs\ can be written as the projection of the only
rotation velocity  along the \los\ \vobs\ $ = $ \vsys\ $+$
\vrot$\cos(\theta)\sin(i)$. Here $\theta$ is the azimuthal angle in
the galactic plane, $i$ the inclination and \vsys\ the systemic velocity of M31.

 In the innermost regions of the disc the largest radial velocities are mostly \textit{not} detected around the (photometric) major axis but at $|Y| > 0$. 
 As a consequence we decided to use the full velocity field  to do
 the tilted-ring model analysis. We have verified that the basic results are not affected by the choice of the opening angle 
around the major axis by comparing with results obtained using a smaller angle (half-sector of 30\degr\ instead of 90\degr).  
As expected,  the parameters derived using the full coverage of the kinematics are better constrained and the rotation curve 
is probed more deeply towards the galaxy centre for the large opening angle than for the small one. The shape 
of the rotation curve is unaffected by the opening angle.

Notice however that a $|\cos(\theta)|$ weight is applied to the data points
during the fitting, giving less weight to pixels close to the minor axis. This is because  
the contribution of \vrot\ to the \los\ velocities is less important along the
minor axis \citep[see also][]{beg89,dbl08}. 

One has to notice that the NE spur ``N2"  and the external arm have been masked for the tilted-ring model analysis because 
their kinematical properties make them not linked to the disc or the other adjacent spur-like structure (see Figs.~\ref{fig:3dview},~\ref{fig:extarm1} and 
\S\ref{discussion-hioutskirts}). 
Leaving them in the velocity map would add large scatter in the results as well as a fail 
of the tilted-ring model at some outer radii.

The location of the dynamical centre and systemic velocity \vsys\ are first fitted
to the velocity field by keeping the disc inclination $i$ and
position angle of the kinematical major axis $P.A.$ fixed at the
photometric values ($\sim 75\degr$ and $\sim 38\degr$ respectively).
 The fitted systemic velocity of M31 is \vsys\ $= -304.5 \pm 6.8$ \kms, whose value is in very good agreement with what is found 
using the integrated spectrum (\S\ref{hicontent}). The location of the dynamical centre is found to 
be offset from the photometric centre by $\sim 0.55$ kpc (in projection). However, the dispersion 
around that location 
is $\sim 0.68$ kpc (in projection), showing that this offset is not really significant.

Then the variation of $i$, $P.A.$ and \vrot\ are measured. The position angle is generally very well constrained with small uncertainties 
(see top panel of Fig.~\ref{fig:rotcur}). 
It can be fixed in a next step to fit the inclination and the rotation curve. The radial profiles of $i$ and $P.A.$ 
sometimes display ring to ring wiggles that may look artificial (Fig.~\ref{fig:rotcur}). This is the reason why 
 the smoothed profiles are used to derive the rotation curve in a final step. 
Model velocity  and residual velocity maps are then generated. The fitting of the parameters and rotation curve 
is repeated until a minimum is found for the average and scatter of the residual map.  

\subsection{Major axis and inclination variations}
\label{sec:tiltedringresults}
\begin{figure}[!ht]
\includegraphics[width=\columnwidth]{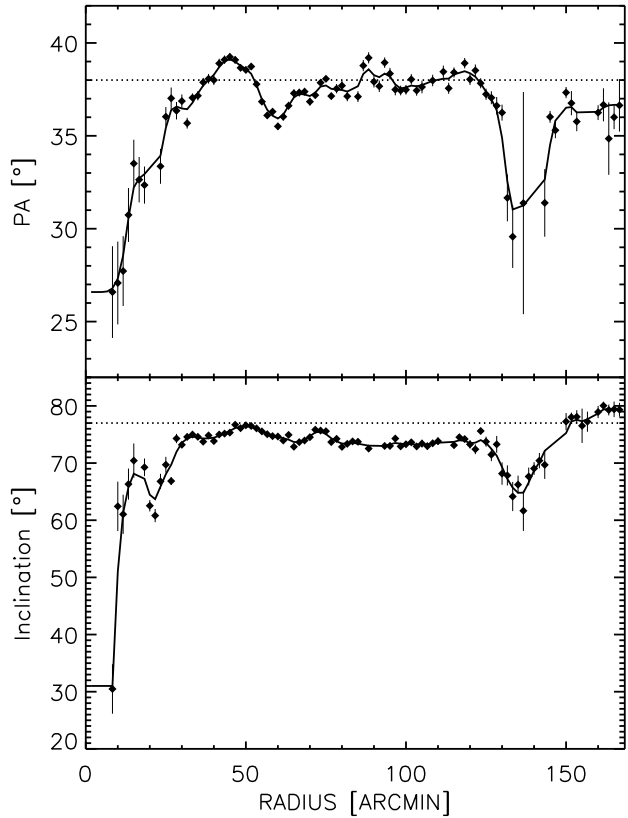}
\caption{\hi\ disc inclination and position angle of the
kinematical major axis as determined from tilted-ring models to the
\hi\ velocity field of Messier 31. Dashed curves represent the smoothed models of $i$ and $P.A.$ used to
 fit the rotation velocities. Horizontal dotted lines show the photometric values.}
\label{fig:rotcur}
\end{figure}

Figure~\ref{fig:rotcur}  displays the
observed and smoothed profiles of $P.A.$ and $i$  and Table~\ref{tab:rotcur} lists the parameters.

We identify five distinct regions in the profiles. First there is a central region $R \lesssim 27\arcmin$ where the inclination
 (position angle) strongly increases as a function of radius by $\sim 40\degr$ (10\degr, respectively). Then a bump is observed 
 between $27\arcmin < R \lesssim 62\arcmin$ for the position  angle which reaches its maximum at $R=45\arcmin$ (\pa $= 39\degr$). Then an extended region between 
 $62\arcmin < R \lesssim 120\arcmin$ where a very small increase of position angle is detected while the inclination remains 
 remarkably constant around 74\degr. Between $120\arcmin < R \lesssim 145\arcmin$ a dip both in the inclination and position angle profiles 
 is detected. The inclination drops down to 61\degr\ and the position angle to 29\degr. Notice that the uncertainties are larger 
 in that radial range. Finally, beyond $R = 145\arcmin$, the position angle remains constant while the inclination is 
 observed to slightly increase towards larger values from 77\degr\ to 80\degr. 
 The average \hi\ disc inclination is $(74.3 \pm 1.1)\degr$ and the average \hi\ disc position angle is $(37.7 \pm 0.9)\degr$, as measured 
 for $27\arcmin < R \la 120\arcmin$ (from 6 kpc to 27 kpc). In that radial range, the kinematical inclination is therefore by 3\degr\ lower than the one 
 derived with optical surface photometry.

\subsection{The HI rotation curve}
\label{sec:hirotcurve}

\begin{figure}[t]
\includegraphics[width=\columnwidth]{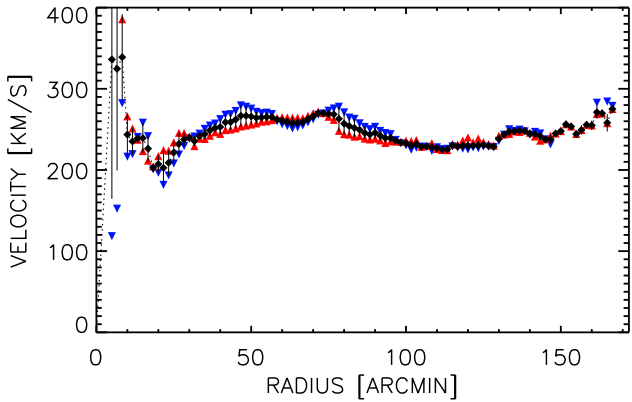}
\caption{\hi\ rotation curve of Messier 31. Filled
diamonds are for both halves of the disc fitted simultaneously while blue
downward/red upward triangles are for the approaching/receding sides
fitted separately (respectively).} 
\label{fig:rc} 
\end{figure}

The \hi\ rotation curve of M31 is shown in Fig.~\ref{fig:rc}. Its velocities are listed in 
Table~\ref{tab:rotcur}. 
The rotation curve is very peaked in the innermost regions, showing velocities up to \vrot $= 340$ \kms. Here, 
the asymmetry between both disc halves is prominent. 
A velocity dip is then observed at $R= 18.3\arcmin$ (4 kpc). The rotation curve is then observed to increase 
up to \vrot $= 267$ \kms\ at $R \sim 47\arcmin$, 
to remain roughly flat between $47\arcmin \la R \la 75\arcmin$ (\vrot$=(264 \pm 4)$ \kms), 
to decrease down to $\sim 230$ \kms, 
to remain flat between $95\arcmin \la R \la 120\arcmin$ (\vrot$=(230 \pm 4)$
\kms) and finally to increase up to \vrot $=275$ \kms\ in the outermost regions.

The uncertainties on the rotation velocities are derived as follows. 
If  gas rotation is made on purely axisymmetric and circular orbits, then rotation velocities are exactly 
the same at opposite sides of the galaxy. If the disc is perturbed, then the axisymmetry is usually 
broken and differences between rotation velocities of the approaching and receding halves exist. 
We thus choose a definition of velocity uncertainties 
 that also takes into account this effect. 
  Uncertainties are indeed defined as the quadratic sum of the formal 1$\sigma$ uncertainty from \textit{rotcur} 
  with the maximum  velocity difference between rotation curves (both disc sides model values derived simultaneously minus 
the approaching or receding side model values derived separately, weighted  
by the number of points in each side). 
With this definition, each uncertainty is  generally larger and more conservative than
 the very small one derived with \textit{rotcur}. 
 It draws a better representation of any departure from axisymmetry.
   As a result  the largest uncertainties are observed in the central regions.
Elsewhere the uncertainties on the rotation velocities are very small, generally lower than 10 \kms. 
It shows the robustness of the method used to select lines to create the velocity field.
Notice that  for some of the annulii beyond $R = 148 \arcmin$ velocities could only be derived 
for the receding half. At those radii, the uncertainty is fixed at the formal $1\sigma$ error provided 
by \textit{rotcur}.
  
\subsection{Comparison with other works}

\subsubsection{The rotation curve}
\label{sec:comprc}
\begin{figure}[ht]
\includegraphics[width=\columnwidth]{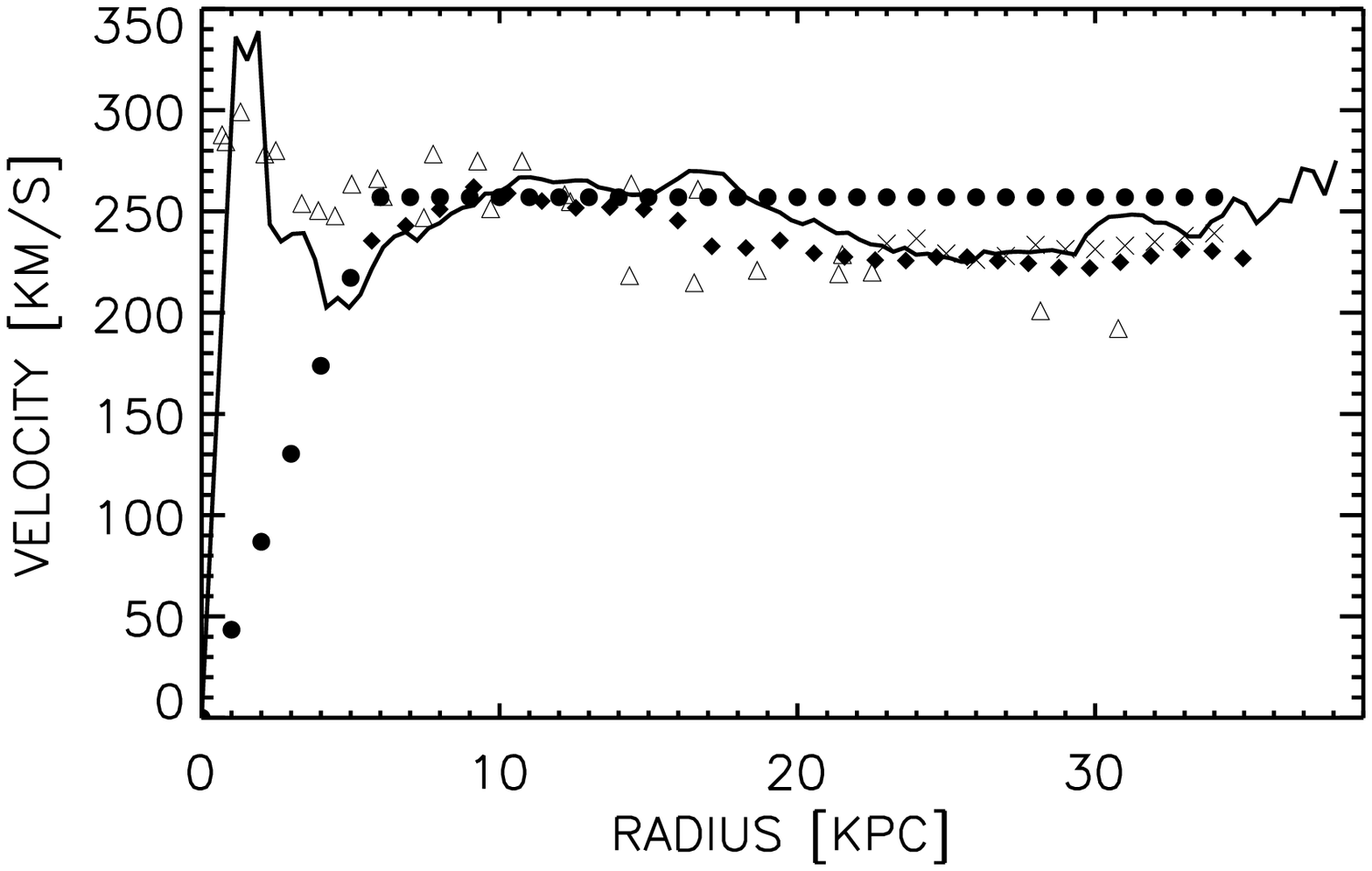} 
\caption{Comparison between \hi\ rotation curves from \citet[][crosses]{new77},
 \citet[][filled circles]{bri84b} \citet[][open triangles]{bra91}, 
 \citet[][filled diamonds]{car06} and our new derived rotation curve (solid line).}
\label{fig:comprc} 
\end{figure}

Figure~\ref{fig:comprc} displays several previous \hi\ rotation curves of M31 
\citep{new77,bri84b,bra91,car06} and our new result. 
The agreement with our previous result \citep{car06} is not good
 between 15 kpc and 23 kpc. This difference can be explained 
 by the fact that \citet{car06} used a single emission line approximation while deriving the velocity field 
 from low  resolution \hi\ data of \citet{unw83}. 
In the external parts  the curves are in better agreement.  
The curve of \citet{car06} is flat at these radii, whose  feature 
is not observed anymore. 
This is likely due to the assumption of constant position angle and inclination in our previous study. 
The agreement with the data points from \citet{new77} looks better. 
Except for $R < 3$ kpc and for some points between $7 < R < 16$ kpc, the agreement with the curve 
of \citet{bra91} is not good. The low velocities he measures around 15 kpc and at large radius (down to \vrot $= 190$ \kms)  
cannot be reproduced from the tilted-ring analysis. The origin of this discrepancy 
may be due to different choices of velocity components to extract the velocity fields. 
The agreement with the model of \citet{bri84b} is good within  $4 < R < 15$ kpc and poor beyond  
19 kpc because the rotation curve is not flat throughout the whole disc.

\subsubsection{The warp of M31}
\label{sec:warpdiscussion}

Results shown in Figure~\ref{fig:rotcur} imply the presence of  two warps  in M31. 
The first warp is located in the central parts ($R \la 6$ kpc) where
 the disc orientation appears less inclined by $\sim$10\degr\ (in average) than 
 the disc. The dip at 4 kpc thus corresponds to perturbed motions located in the inner warped disc. 
 The nuclear region  ($R < 2$ kpc) is even more perturbed as 
 it behaves like a ``warp in the warp". Here rotation velocities,  
inclination and position angles are very different than anywhere else in the disc. 
 Gas is observed to be very close to face-on. 
 The very large uncertainties on the rotation velocities show that  noncircular 
motions may be very important here. 
 A second warp is detected in the outer parts ($R > 27$ kpc), 
 where the disc is surprisingly oriented like in its inner warped part (\pa\ and inclination)
 and then becomes more and more inclined at the largest radii. The fact that \hi\ gas is observed to 
 start rotating faster where the outer warp appears is also probably a consequence of the inferred 
 perturbed orbits. Notice however that the symmetry is excellent between both halves whereas 
 it is not the case in the inner warp.

This is not the first time that warps are evidenced in M31. One can mention results by 
 \citet{cia88} and \citet{bra91} for  the inner regions, and \citet{new77}, \citet{hen79} or \citet{bri84b} for
 the outer regions.

\begin{itemize}
\item  \citet{bra91} showed that the \hi\ distribution is tilted by $15\degr$ 
in the inner 10\arcmin\ (2.3 kpc). We confirm the presence of the inner
 \hi\ warp in the DRAO observations. It is actually observed to extend up to $R = 6$ kpc for
  a maximum tilting angle of $\sim 15\degr$ and a maximum twisting angle of $\sim 7\degr$ with respect 
  to the average disc inclination and position angles as given above. 
  If the perturbation in the nuclear region is genuine, then the warp is 
  even more prominent, with maximum tilting and twisting angles of $42\degr$ and $10\degr$.

\item The ionized gas distribution is more  circular in the bulge, implying a more face-on disc \citep{cia88}.
 Our measurement confirms this trend.

\item
 \citet{new77} derived warp parameters in the outer disc ($R > 28.5$ kpc, or 125\arcmin), with a decreasing 
position angle (down to 30\degr) and an increasing inclination (up to 83\degr). 
The shape, amplitude and locaion of the outer warp 
detected in the DRAO data are in agreement with their early result, though the 
inclination drops before becoming larger. 

\item
  Another   modelization was done by \citet{hen79}, 
 in which the warp starts at 18.2 kpc (80\arcmin) with an increasing inclination in step 
 of 0.75\degr\ per kpc, and a warped region rotated by 10\degr\ with respect to the central disc position 
 angle. This result is not totally consistent with our new result.
  
\item In the model of \citet{bri84b}, the disc starts to warp
 at $\sim 18$ kpc, with a maximum warp angle of 15\degr. 
We detect a gradient of inclination in agreement with that value, though it is seen to occur 
at larger radius. 

\end{itemize}

According to \citet{bri84b}, 39\% of the total \hi\ mass resides in the warped part of the disc. 
This number has to be compared with the   gas mass of the main component 
integrated within $R > 27$ kpc ($0.14 \times 10^9$ \msol) added to the total mass of all other spectral
components than the main one ($1.74 \times 10^9$ \msol). 
It corresponds to 44\% of the total 
\hi\ mass. This fraction is comparable with the prediction of \citet{bri84b}, though being slightly higher.

 \section{Origin(s) of the other spectral components}
 \label{sec:otheremissionline}
 
 No kinematical analysis is presented for the other spectral components than the main disc component. 
Only a brief discussion on their possible origin(s) is proposed here.
So many \hi\ peaks like those observed in M31 are rarely evidenced in extragalactic sources.
The sum of all of the integrated emission other than the main component represents 41\% of the total \hi\ mass. 
A significant fraction of it is due to the overlap emission from the warp whose signature is more prominent 
at large $Y$ than along the major axis due to the inclination of the disc.  Another part is due 
to an additional gas component whose origin(s) can be manifold. 

A first origin could be structures lying in the disc itself, like e.g. unresolved spiral arms by projection effects. 
Indeed, the 3$^{\rm rd}$, 4$^{\rm th}$ and 5$^{\rm th}$ maps of 
Figure~\ref{fig:hivf1} show that projection effects may play an important role in creating multiple 
components due to the presence of more pixels at $|Y| > 0$  than along the major axis. 
Moreover the projected spatial distribution of the gas in the multiple peaks is principally concentrated 
within high surface density regions, like the \hi\ ``ring" around 13 kpc and along the spiral arms.
Notice here that more components are detected in the north-western front side of the galaxy than on the far side. 
Another internal origin of multiple components could be caused by expanding gas shells 
induced by stellar winds in star forming regions.  Such a phenomenon has already been seen in other galaxies \citep[e.g.][]{hun01}.  
Several spectra of M31 show more or less symmetric peaks centered about the main component and
could point out gas outflow in high density regions of the M31 disc.  

Other origins could be extraplanar gas in the form of e.g.~a lagging halo or high velocity clouds.
In recent deep \hi\ observations a lagging halo corresponds to the thick gaseous layer that is observed to rotate more slowly 
than the host equatorial thin \hi\ disc of spiral galaxies \citep{fra01,fra02,bar05,oos07}. The halo emission
which is often referred to as an ``anomalous" emission  has a mass that can reach $\sim 30\%$ of the total gas mass \citep{oos07}.
It is thought to be gas infalling onto the disc, mainly due to a galactic fountain mechanism  but also to the accretion from 
the intergalactic medium \citep{fra06,fra08}. Models of gas in the halo show it has a larger velocity dispersion than gas in the cold disc 
\citep{oos07}. The DRAO datacube exhibit many \hi\ peaks whose radial velocity is observed to be closer to the systemic velocity than the main \hi\ disc component 
and whose velocity dispersion is larger than the main disc component. It implies that M31 could host a lagging halo as well. 
 
 High or intermediate velocity clouds (HIVC) similar to those orbiting around the Milky Way \citep{wak99} 
 are observed in a few galaxies  among which M31 is a very good candidate \citep{thi04,wes05}. 
 M31 HIVC detected by \citet{wes05} have a typical mass $\propto 10^5$ \msol\ for a size of $\sim$ 1 kpc.
 Similar high or intermediate velocity clouds could be detected in the DRAO field-of-view.  
 Indeed the minimum detectable column density of the data  
 corresponds to $1.6\times 10^4$ \msol\ per spatial resolution element (synthesized beam size). 

 One finally notices that no obvious forbidden velocities are detected in the high resolution datacube of M31 
 at a 3$\sigma$ detection level. 
 Forbidden velocity clouds (FVC) correspond to apparent counterrotating gas, like 
  in  NGC 2403, NGC 891 or NGC 6946  \citep{fra02,oos07,boo08}. The presence of FVC could 
  point out ongoing accretion onto a host disc.
 As a consequence if any \hi\ gas accretion responsible of part of the multiple lines
 is occuring in M31, it does not seem to be done from apparent counterrotating material.   
 A more careful search of HIVC or FVC in our datacube will require to 
 filter the high resolution datacube in the spatial and spectral dimensions to increase 
 the signal-to-noise ratio.

\section{The \hi\ outskirts}
\label{discussion-hioutskirts}

\subsection{The NE and SW spurs}
\begin{figure*}[!t]
\begin{center}
\includegraphics[width=\textwidth]{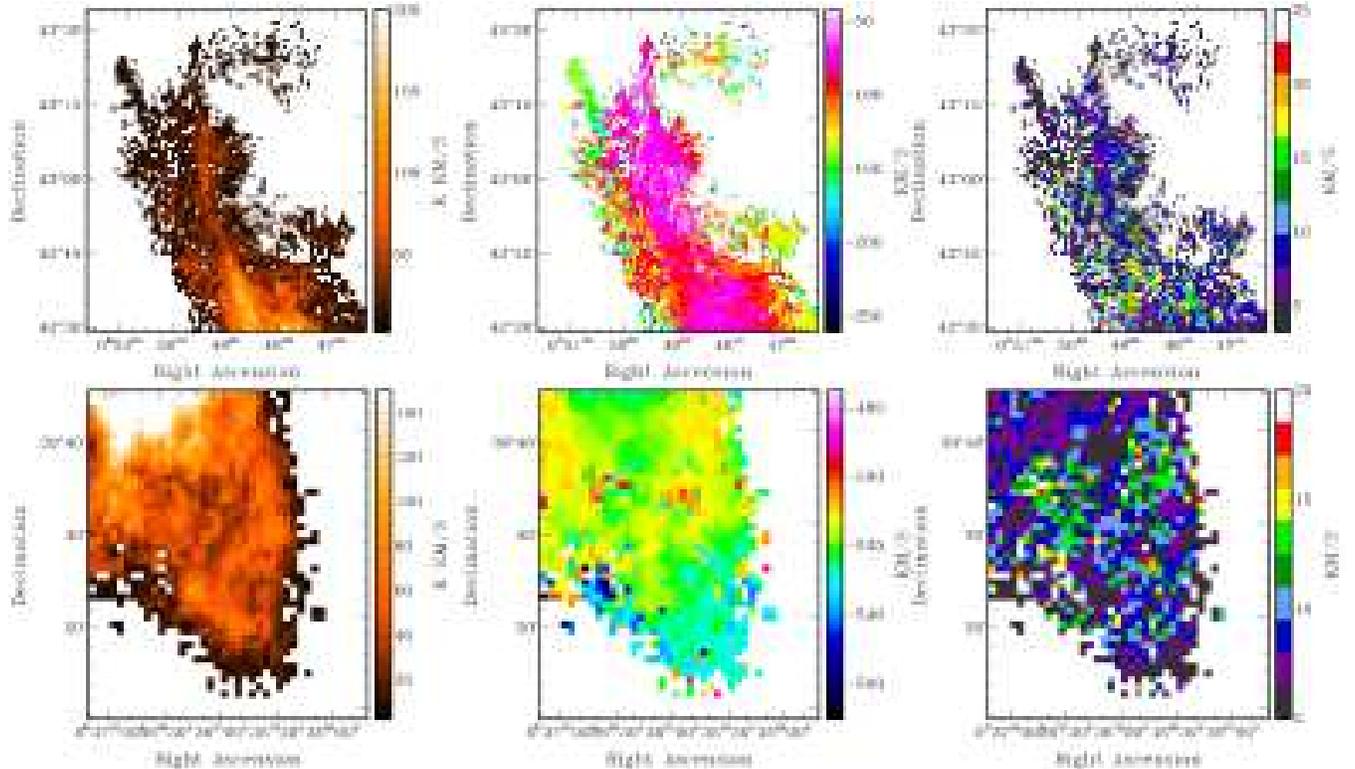}
\caption{\hi\ emission, velocity and velocity dispersion maps zoomed in the region of the NE (top panel) 
and SW (bottom) structures.}
\label{fig:neregion}
\end{center}
\end{figure*}

Neutral gas in  M31 is detected out to $\sim 34.1$ kpc  in projection ($\sim 36.3$ kpc) in the 
SW approaching side (NE receding side respectively). 
Two extended thin spurs are observed at the north-eastern extremity of the disc while gas  seems 
more confined in its south-western extremity.  
These structures are connected to the spiral arms as seen in Figs.~\ref{fig:hitot},~\ref{fig:3dview} and ~\ref{fig:neregion}. 

 A column density of  $6.2\times 10^{20}\ \rm cm^{-2}$ is reached in the south-western \hi\ extension. 
 A velocity gradient is detected from $-515$ \kms\  to $-530$ \kms\ for a length of 
 $\sim 30\arcmin$ (6.8 kpc), in apparent continuity of the disc kinematics. 
 The velocity dispersion of the \hi\ peaks is $12$ \kms\ in its northern part while and lower values  
 are seen in its southern part. 
 
The NE spur (N1) is oriented North-South and has a length of $\sim 40\arcmin$ (9.1 kpc). Its 
kinematics is in continuity with that of the disc. A velocity gradient from $-85$ \kms\ 
to $-65$ \kms\ is detected along it while the velocity dispersion of the \hi\ lines remains constant ($\sim 8-10$ \kms). The 
highest \hi\ density column observed in it is $4.8\times 10^{20}\ \rm cm^{-2}$. 
A small ``hook-like"  structure at $\alpha_{2000} = \rm 00^h 49^m 11.86^s, \delta_{2000} = +43^d 06\arcmin 24\arcsec$ is 
observed as part of the spur. 
The other spur-like structure (N2) is shorter ($\sim 30\arcmin$). 
A velocity gradient is also observed along it, from $-160$ \kms\ to $-140$ \kms.
 Its velocity dispersion is very low ($\sim 5$ \kms) and is also seen to be constant.  
  An average column density of $6\times 10^{19}\ \rm cm^{-2}$  is derived for it.
 Notice that the three \hi\ extensions have very few pixels with multiple spectral components. 

\begin{figure}[!t]
\begin{center}
\includegraphics[width=\columnwidth]{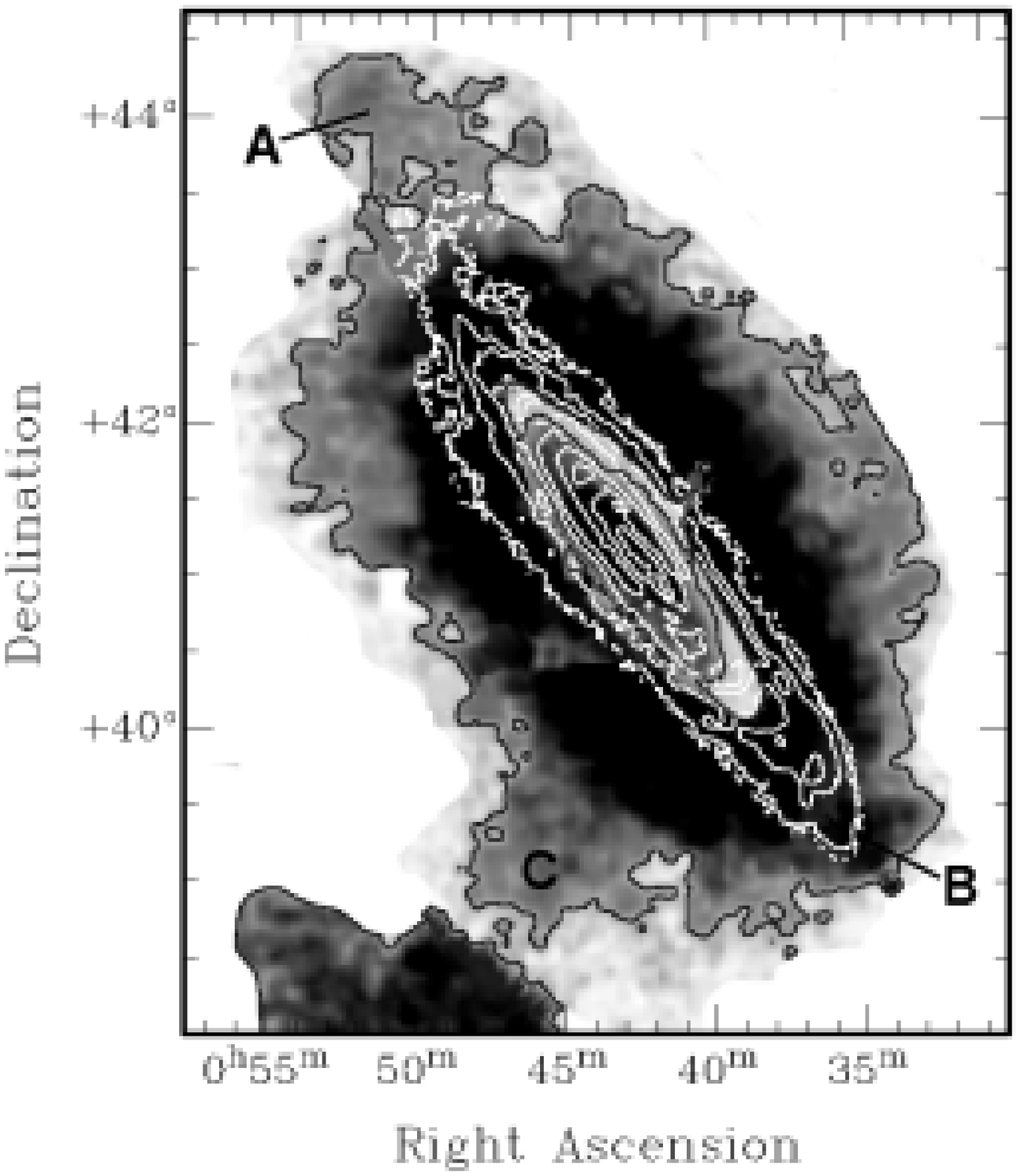}
\caption{Comparison of the \hi\ distribution (white contours) with the faint extended stellar disc of M31 from \citet[][grey-scale image
 and grey contours adapted from their Fig. 50.]{iba07}. The letter A refers to the North-East stellar clump, B to the G1 clump and C to the Giant Stream, following the nomenclature 
 of  \citet{iba07}. 
\hi\ contours are for column densities of $5\times 10^{19}$, $1\times 10^{20}$, $4\times 10^{20}$, $9\times 10^{20}$ and $2\times 10^{21}$
 cm$^{-2}$.  }
\label{fig:deepstellar}
\end{center}
\end{figure}

 Are these structures real?  What could be their origin(s)? 
 One has to notice first that these extensions are genuine structures of the disc 
 because they are also observed in the WSRT image \citep{bra09} and  
  they cannot be mistaken with residuals from a bad  subraction of  MW \hi\ due to their kinematics. 
 Then the kinematical properties of the two NE spurs strongly differ from each other. 
 We argue they are part of two different \hi\ structures, though they are located in the same area 
 of the field-of-view.  
  On one hand a close inspection of the datacube points out that the spur N2 seems to be a kinematical
 extension of the external spiral arm of M31 (Fig.~\ref{fig:3dview}). 
  On another hand the brightness and kinematical properties of the spur N1 and the SW extension 
  are quite similar. In particular they are responsible of the velocity rise that occurs at 
  the outermost radii of the rotation curve. 
  What is also remarkable with them is that they 
  both overlay or point towards two diffuse stellar clumps, as seen in Fig.~\ref{fig:deepstellar}. 
  This image displays the \hi\ column density contours superimposed onto the stellar distribution, as measured from deep photometry \citep{iba01,iba05,iba07}. 
The SW extension is coincident with the G1 clump while the NE spur   
coincides with part of the northeast clump. 
According to \cite{iba05} stars in the G1 clump rotates by 68 \kms\ faster than the disc,  
which is considered to rotate at $\sim 230$ \kms\ in their model.
The true rotation velocity in the south-western extension is not observed constant at large radii, but increasing. This increase 
implies a lagging velocity of up to $45$ \kms\ larger than the velocity of $\sim 230$ \kms. The lagging 
velocity of gas is thus lower than that of stars. 
The spur N1 to the North-East is observed to rotate faster than \vrot\ $=230$
\kms\ as well by a similar amplitude than in the southern spur,
 while the NE stellar clump is observed to rotate slower. 
Hence gas and stars follow the same velocity trend in the SW region of the disc, while it does not seem to be the case in the 
NE region. Notice however that stellar velocities are less certain in the NE clump because of the contamination by MW stars.
It nevertheless indicates a strong relationship between stars and gas at the outskirts of M31, at least in the SW region. 

The star formation history and metallicities of the northeast and G1 overdensities 
led \citet{far07} and \citet{ric08} to propose they could be material initially formed into the disc 
that have been stripped 
by tidal effects. 
The kinematics of the \hi\ spurs imply that they are bound to the disc. Though they seem to rotate faster than gas, 
it is likely that stars in the external clumps are also bound  to the disc of M31. 
A fit to the velocity field of a radial motion $v_{\rm rad}\sin(i)\sin(\theta)$ in addition to the rotational one gives 
 an  average value $\bar{v}_{\rm rad} = 6.2$ \kms\ with a standard deviation of 7.4 \kms\ 
 for $R > 150\arcmin$ ($R > 35$ kpc).
 
 By assuming that the spiral arms are trailing in the disc and by looking at the distribution of dust lanes in optical images of M31, 
the front side of the galaxy is to the NW of the major axis and the rotation is done counterclockwise. 
Therefore, at first order outflow motions  are detected in the outskirts of M31 because of the derived positive radial velocities.  
At second order these radial motions could be mistaken by 
 the presence of either vertical motions to the galaxy plane, which would thus be of the order of 35 \kms\   
 for the observed  inclination ($80\degr$, Fig.~\ref{fig:rotcur}), or elliptical streaming in  e.g. a 
 $m=2$ perturbing potential. However 
 that later hypothesis is less uncertain here because no evident spiral structure is seen at those 
 radii but rather thin ``filamentary" \hi\ spurs. The reality is probably a combination of these hypotheses 
 (pure outflow, z-motions and elliptical
 streaming). $\bar{v}_{\rm rad} \sim 6$ \kms\ is likely an  upper limit of any possible outflowing motions.  
 Such radial motions are consistent with the scenario proposed by \citet{far07} or \citet{ric08}. 
Indeed, one expect outflow motions in the disc outskirts if tidal stripping is occuring. 
N-body models would be helpful to firmly validate the detection of ongoing outflow in the outskirts of M31.

\subsection{The external arm}
\begin{figure}[!h]
\includegraphics[width=\columnwidth]{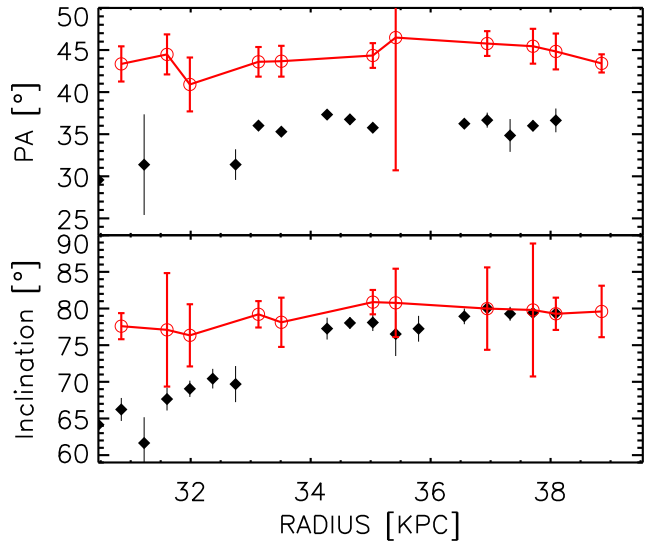}
\caption{\hi\ inclination and position angle of the external arm (open symbols). Filled symbols show the 
orientation of the disc parameters from Fig.~\ref{fig:rotcur}.}
\label{fig:extarm1}
\end{figure}

The external arm is another new perturbed structure in M31. It has many properties different from other disc structures.
It has no identifiable morphological and kinematical symmetric counterpart\footnote{The only noteworthy \hi\ feature 
which could be a counterpart on the other edge of the disc is located around  
$\alpha_{2000} = \rm 00^h 39^m 49.80^s, \delta_{2000} = +41^d 13\arcmin 39\arcsec$. However it is not an 
extended structure like the external arm.}
 with respect to the galactic centre. 
Its kinematics significantly differs from radial velocities within the disc (Fig.~\ref{fig:3dview}). 
It does not contain so many pixels with multiple \hi\ peaks. 
It is one of the faintest structures detected in the field-of-view with a brightness comparable to 
interarm regions. The total \hi\ mass attached to this structure is $1.1 \times 10^8$ \msol. Its apparent length is 2.33\degr\ or 32 kpc.
The kinematical difference with the disc is caused by different orientation angles. Indeed its position angle 
and inclination are  $\sim$ 10\degr\ larger than those of the disc (Fig.~\ref{fig:extarm1}), with the exception of the region $R > 34$ kpc 
where both inclinations become very similar. The rotation velocities of the external arm 
are comparable with the disc rotation curve. No obvious stellar counterpart to the gaseous arm can be 
identified in the optical images of \citet{iba05,iba07} because of their too high contrast.

Here again one may question about its origin. 
Internally driven perturbations are hardly possible because one would expect the creation of 
axisymmetric or bisymmetric features in the gas distribution (e.g. a bar, a ring or two spiral arms). 
The apparent isolated nature of the arm on one half of the disc and not on the other side 
rules out this hypothesis and points out to an external perturbation for its origin. 
It includes spiral arm triggering  by tidal effects exerted by a companion 
or even gas accreted from a companion. This would be more consistent given the perturbed nature of the stellar distribution 
at large radius caused by tidal interactions with low mass satellites. 
Indeed the stellar halo of M31 is filled of stellar residuals from past   interactions with smaller galaxies. 
M31 has currently two nearby bright satellites, M32 and NGC 205 at  projected distances of $\sim 25$ kpc and $\sim 37$ kpc to the M31 nucleus. 
Both are expected to have interacted with M31. 
In a gas accretion scenario, one would need an efficient tidal stripping because both companions are devoid of gas. 
\citet{blo06} simulated a head-on collision with  M32 that could have occured 210 million years ago and have generated 
the ring-like morphology in the distribution of dust and gas. \citet{how08} simulated an interaction with NGC 205 
and found radial orbits at high velocity (up to 500 \kms) for the small companion. 
According to the possible orbits from the models, we think the most likely impactor that generated the external arm
 could be NGC 205. The location of the perturbed external arm indeed coincides with the path of NGC 205 in the 
 apparent far side of the galaxy \citep[see Fig. 19 of][]{how08} while the M32 trajectory leads the compact elliptical 
 close to the M31 centre along a polar orbit. 
  Here again  more detailed numerical simulations are needed to investigate the different possibilities 
 and find the origin of this perturbation.
  
\section{Mass distribution analysis}
\label{massmodels}

The mass distribution modeling of the galaxy is done by decomposing
the total gravitational potential into a supermassive black hole, 
a luminous baryonic component and a dark matter component. 
It consists in fitting a dynamical model for the dark matter component to the
observed rotation curve, taking into account the  baryonic
rotation curves. Two or three free parameters are fitted for the 
halo in addition to the stellar mass-to-light ratios for the bulge
and galactic disc. These latter can be fixed under some assumptions
described below.

\subsection{The central supermassive black hole}
Messier 31 is known to have a central supermassive black hole which
mass can now be robustly constrained. The black
hole contribution to the rotation curve remains negligible through
the disc because of the point mass nature of its gravitational
potential. A mass of $1.4 \times 10^8$ \msol\ was derived using
Hubble Space Telescope STIS spectroscopy of the nucleus
\citep{ben05} and is used in our study.

\subsection{The luminous baryonic matter}
The contribution of the luminous baryons to the rotation curve
consists in stellar and gaseous components.

\subsubsection{The stellar potential}
\label{sec:stelpotential} The gravitational potential of the stars
is decomposed into contributions from a bulge and a disc.  These ones 
are derived from the $R-$band surface brightness profile of
M31 \citep{wal87,wal88}.

\begin{figure}[!h]
\includegraphics[width=\columnwidth]{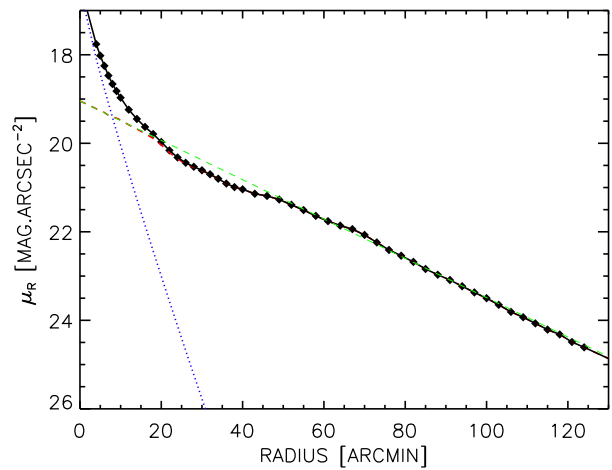}
\caption{$R-$band surface brightness profile  of Messier 31. Observations are from \citet{wal87}. 
Filled symbols are the observations, a red dashed line is the adopted contribution of the disc to the surface brightness, 
a blue dotted line the adopted contribution of the bulge and a solid line the sum of both components. A green dashed line 
shows the fit of the exponential disc.}
\label{fig:rbanbrightness}
\end{figure}

 A bulge-disc decomposition of the surface
brightness profile is done using an exponential disc contribution
for the stellar disc plus a de Vaucouleurs $R^{1/4}$ law
\citep{dev48} for the bulge contribution. However, the de Vaucouleurs
law gives a bad result for the bulge contribution to the
surface brightness profile because it significantly overestimates the intensity 
in the inner parts of the galaxy.
A more robust  bulge-disc decomposition is done
using a generalized S\'ersic $R^{1/n}$ model for the bulge light
\citep{ser63,ser68} plus an exponential disc. In the S\'ersic model,
the intensity profile is written as
\begin{equation}
I(R)=I_e \exp\left\{ -b_n\left[ (R/R_e) ^{1/n} -1\right] \right\},
\label{eqsersic}
\end{equation}
in which $I_e$ is the intensity at the effective radius $R_e$ defined as
the radius that contains half of the total light and $n$ a dimensionless index that
measures the ``curvature" of the luminosity profile. $b_n$ is a function of
$n$ and is defined as $b_n = 1.9992n - 0.3271$ for $0.5 < n < 10$ \citep{cap89}.
The de Vaucouleurs law is thus a specific case of the S\'ersic law for $n=4$.

The stellar disc intensity profile is fitted with the exponential disc formula
\citep[see e.g.][]{fre70}
\begin{equation}
I(R) = I_0\ \exp \left(- R/R_d\right)
\label{eqexpo}
\end{equation}
where $I_0$ is the  intensity at $R=0$ and $R_d$ the scale-length of the disc.

The two components were fitted simultaneously to the brightness
profile (Fig.~\ref{fig:rbanbrightness}).  The fitted bulge
parameters are $R_e = (1.3 \pm 0.1)$ kpc, $\mu _e = (18.6 \pm
0.1)$ mag arcsec$^{-2}$ (not corrected for foreground and internal dust attenuation) and $n = 1.1 \pm 0.1$. 
The total apparent magnitude of a bulge for the S\'ersic law is given by  :
\begin{equation}
 m = \mu_e - 5\log R_e -2.5 \log \left[ 2\pi n\frac{{\rm e}^{b_n}}{(b_n)^{2n}}\Gamma (2n) \right]
\end{equation}
\citep{gra05}, which gives a total uncorrected apparent magnitude of $m_{\rm Bulge} \sim 3.25$ mag for M31. 

The fitted disc parameters are $R_d = (5.6  \pm 0.1)$ kpc and a central
surface brightness $\mu_0 =  (19.0 \pm 0.1) $ mag arcsec$^{-2}$ (not corrected for foreground and internal dust attenuation nor inclination effects). 
This disc scale-length compares well with
values fitted with the same dataset \citep{wal87,wal88,gee06}. 
 As seen in Fig.~\ref{fig:rbanbrightness}, 
the adopted disc brightness profile is nevertheless not the fitted one, but the result of the subtraction 
of the fitted bulge contribution to the observed profile.  
This allows us to keep the stellar surface brightness as close as possible to the observed profile.
Moreover, it enables us to keep the intensity variations of the disc that could reproduce   wiggles in the rotation curve. 
The surface brightness profiles are then corrected from Galactic extinction \citep[following][]{sch98} and for
inclination effects (only for the disc profile) and for internal extinction as described in \S\ref{sec:stelmasstolightratio}.

\subsubsection{The mass-to-light ratios of the stellar component}
\label{sec:stelmasstolightratio} Two free parameters of the mass
models are the mass-to-light ratios of the bulge \mlb\ and of the
galactic disc \mld, which are considered constant as a function of radius for the remainder. Models presented here also
enable these parameters to be fixed. A  relative approximation for them 
can be deduced from stellar population synthesis (SSP) models
\citep[e.g.][]{bel03}, provided that a color is given. 

The bulge color is $B-R = 1.75$ mag \citep{wal88}, which gives a color of $B-R \sim 1.27$ 
after correction for foreground and internal dust extinction effects. 
Here we have applied the reddening law for M31 derived by \citet{bar00}, using a mean reddening factor 
$E_{B-V} = 0.24$ mag \citep{wil01}. This reddening law is relatively comparable with the one measured
 for the Milky Way.
The corrected bulge mass-to-light ratio is \mlb\ $\sim 2.2$ \msol/\lsol, following the prescriptions of \citet{bel03}. 
These authors indicate a r.m.s. scatter of the order of 25\% on the mass-to-light ratios.
An independent constraint of the bulge mass-to-light ratio can be obtained directly from measurements of the 
bulge effective stellar velocity dispersion, with the assumption that the dark halo has a negligible contribution
to the velocity dispersion. The estimated dynamical mass of the bulge is 
$\M_{\rm Bulge} = 1.9 \times 10^{10}$ \msol\ \citep{mar03}, which value becomes $\M_{\rm Bulge} = 2.4 \times 10^{10}$ \msol\ 
after scaling to our adopted bulge effective radius. It implies a corrected dynamical mass-to-light ratio
ratio of  \mlb\ $\sim 0.8$ \msol/\lsol. This is about 2.8 times lower than the ratio derived from SSP models. 

The average disc color is $B-R \sim 1.6$ mag, as derived from \citet{wal87}, 
giving a corrected color $B-R \sim 1.1$ mag and a corrected disc mass-to-light ratio \mld\ $\sim 1.7$ \msol/\lsol. 
When no correction from internal extinction is applied to the $B-R$ colors, 
SSP models imply \mlb\ $\sim 4.0$ and \mld\ $\sim 3.2$.
Our values of corrected and uncorrected \mld\ are thus in  good agreement with those
 estimated by \citet{wid05} or \citet{gee06}. 

Applying the same foreground and internal corrections as above \citep[using $A_B = 0.67$ mag,][]{dev91}, the 
total bulge luminosity of M31 is $\rm L_{Bulge} \sim 2.9 \times 10^{10}$ \lsol\  in the $R-$band, which 
translates into a total corrected bulge mass of $\rm \M_{\rm Bulge} \sim 6.4 \times 10^{10}$  \msol\ 
 for the adopted \mlb\ $= 2.2$. 
The deduced total luminosity of the disc (also corrected for inclination) is 
$\rm L_{Disc} \sim 4.2 \times 10^{10}$ \lsol, corresponding to a
total corrected stellar disc mass of $\M_{\rm Disc} \sim 7.1 \times 10^{10}$  \msol\ 
for the adopted \mld\ $= 1.7$.
The total inferred stellar mass is $\M_{\star} \sim 1.35\times 10^{11}$ \msol. 
SSP models thus infer disc and bulge masses comparable with each other, which is somewhat unexpected for a
 high surface brightness galaxy \citep{cou99}. If one uses the dynamical value \mlb\ $= 0.8$ 
instead of the SSP value, then the total stellar mass becomes
$\M_{\star} \sim 9.5\times 10^{10}$ \msol, making the bulge-to-disc mass ratio falling down to 34\%, 
which seems more realistic. 

\subsubsection{Infrared surface photometry}
Determining mass-to-light ratios for galaxies in the 3.6 \micron\ band from stellar synthesis 
models is not an easy task because no straightforward recipe exists at this wavelength, 
contrary to optical bands.

We have attempted to derive stellar mass-to-light ratios and masses
 from observations performed with the \textit{SPITZER}-Infrared Array Camera at 3.6
\micron\ \citep{bar06}. However no realistic stellar masses and mass-to-light ratios 
could be estimated, using either the formalism developed in \citet{dbl08} that derives  
mass-to-light ratios  from the Large Galaxy Atlas of the 
 Two Micron All Sky Survey (2MASS) $J-K$ bulge and disc colors, 
or the average values of bulge and disc mass-to-light ratios inferred for seven barred and unbarred 
 Sab, Sb and Sbc spirals  from the THINGS galaxy sample of \citet{dbl08}.
  Such values both lead to either a too large total stellar mass or
  a too massive bulge compared with the disc.  As a consequence mass distribution  models have been 
  fitted using the only $R-$band photometry  for clarity and simplicity reasons.
   
\subsubsection{The gaseous disc potential}
\label{sec:gasdens} 

The contribution to the rotation curve from the
gaseous component is derived from the \hh\ and \hi\ surface densities.
The \hh\ surface density is taken from the CO gas survey of \citet{dam93}, 
after scaling to our adopted distance. The total molecular 
mass is $\sim 3.5\times 10^8$ \msol.
The \hi\ surface density profile is determined by averaging the total 
\hi\ emission over elliptical rings which orientations are given by
the tilted-ring model described in section~\ref{sec:tiltedring}. 

\begin{figure}[!t]
\includegraphics[width=\columnwidth]{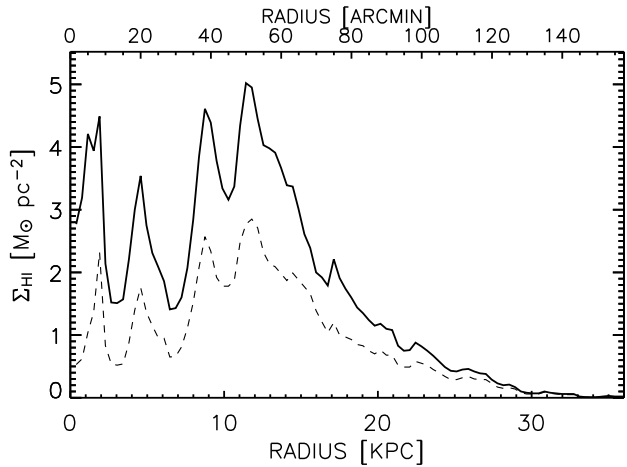}
\caption{\hi\ surface density profile of Messier 31. A full line
is for the total \hi\ emission, a dashed line for the emission of the main component.}
\label{fig:gasdens} 
\end{figure}

The \hi\ surface density profile is displayed in
Figure~\ref{fig:gasdens}, for the total \hi\ distribution and for the principal 
emission component (the one that served to build the velocity field). 
The profile of the total \hi\ distribution is listed in Tab.~\ref{tab:rotcur}. 
It has been derived using orientation parameters of Fig.~\ref{fig:rotcur} and Tab.~\ref{tab:rotcur}.
 No correction due to the overlap from the warp emission has been applied to the 
 density profile because no 3D model of the datacube has been attempted in this article. 
 It has no  consequences on the following result because of the very low induced amplitudes 
 of the atomic gas rotation velocities.

The majority of \hi\ gas is located between $R = 7$ kpc and $R=18$ kpc, which  corresponds to the location of 
the brightest ``rings". Peaks are also observed in the inner region, in agreement with the 
locations of other inner ring-like structures. 
The rotation curve of the atomic gas component is then derived after scaling the
 density profile by a factor of $\sim 1.3$ in order to take  into account the helium contribution. 
The contribution of the atomic gas to the overall
 rotation curve is very low and does not exceed $\sim 70$ \kms. 
As the mass of the molecular gas is only $\sim 8$\% that of the total 
atomic gas, its contribution to the rotation curve is also 
very low ($\sim 15$ \kms\ at maximum) and comparable with the velocities due to the 
 black hole at the largest distances. 
 The mass-to-light ratios of the molecular and atomic components are kept fixed during the fittings 
 ($\Upsilon_{\rm gas} = 1$).

\subsection{The dark matter halo potential}
\label{sec:dm} Three  different halos are fitted in this study. A first
model is associated with a pseudo-isothermal sphere while two other  
models are cosmological halos as their density profile is derived from Cold Dark Matter (CDM)
numerical simulations. No attempts to
fit the mass distribution in triaxial models were done as all models consider a 
spherical halo. Also, no attempt to model an adiabatic contraction of the dark
 halo in its central parts was done  \citep{dut05}.

\subsubsection{Navarro-Frenk-White halo}
\citet{nfw96,nfw97}  were among the first ones to propose a formalism that was fitted to results from   
numerical simulations done in the framework of the CDM theory. 
They found a halo shape that is independent of the halo mass, which 
is why this halo is often referred to as the ``universal" halo. The mass density profile is steep (cuspy)  as it scales 
with $R^{-1}$ at low radius and is written as
\begin{equation}
\rho_{\rm nfw}(R)=\frac{\delta_c\ \rho_{\rm crit}}{(R/R_s)(1 +(R/R_s)^2)},
\label{eqnfw}
\end{equation}
where $\rho_{\rm crit} =  3 H^2_0/(8\pi G)$ is the critical density for closure of the Universe, $\delta_c$ 
represents a characteristic density contrast and $R_s$ a scale radius. The 
circular velocity profile corresponding to this halo  allows to fit 
two parameters, $V_{200}$, the velocity at a the virial radius $R_{200}$, and $c=R_{200}/R_s$, a concentration parameter of the halo, 
to the observed rotation curve, and is written as :
\begin{equation}
V (R)=V_{200}\sqrt{\frac{1}{x}\frac{\ln(1+cx)-cx/(1+cx)}{\ln(1+c)-c/(1+c)}}.
\label{eqnfw}
\end{equation}
In this equation, $x=R/R_{200}$. This model is referred to as the ``NFW" model or NFW halo hereafter.

\subsubsection{Einasto halo}
\citet{mer05,mer06} used more recent $\Lambda$CDM numerical simulations to model the mass density profile of dark halos
and found that they can be fitted with a three-parameter model. The density is written as
\begin{equation}
\rho_{\rm ein}(R)=\rho_{\rm e} \exp\left\{ -d_n\left[ (R/R_{\rm e})^{1/n} -1\right] \right\}.
\label{eqeinastodens}
\end{equation}

This relation is very similar to the expression of the luminosity
profile for elliptical galaxies and bulges of galaxies
(Eq.~\ref{eqsersic}), with the difference that the S\'ersic formula
applies to the (projected) surface distribution of light from
galaxies whereas Eq.~\ref{eqeinastodens} applies to the spatial
distribution of the halo mass. Like in Eq.~\ref{eqsersic}, $n$ is a dimensionless parameter which 
measures the ``curvature" of the density profile, $d_n$ is a
function of $n$ and can be approximated by $d_n \approx 3n - 1/3 +
0.0079/n$ \citep{mer06}, provided that $n \gtrsim 0.5$. Here,
$\rho_{\rm e}$ is the density at the effective radius $R_{\rm e}$
defined as the radius of a sphere in which half of the total
halo mass is contained.

The mass profile of the Einasto halo \citep{car05, mam05, mer06} is written as
\begin{equation}
M_{\rm ein}(R) = 4\pi n R_{\rm e}^{3} \rho_{\rm e} {\rm e}^{d_n} {d_n}^{-3n} \gamma (3n,x).
\label{eqeinastomass}
\end{equation}

Here $\gamma(3n,x)$ is the incomplete gamma function defined as
\begin{equation}
\gamma (3n,x)=\int ^{x}_{0} {\rm e}^{-t}t^{3n-1} {\rm d}t.
\end{equation}

\citet{mer06} and \citet{gra06} refer to this model as the Einasto halo due to the early works made by \citet{ein65,ein68,ein69} 
and \citet{ein89} on the light and mass distributions of galaxies, independently from those of \citet{ser68}. 
Simulated empirical galaxy-sized halos have typical values $n = 5-7$ \citep[Tab.1 of][]{mer06},
$\log(\rho_e) \sim -5.5$ (\msol\ pc$^{-3}$) and $r_e$ of several hundreds of kpc. 
Hereafter, this model is referred to as model ``EIN" or Einasto halo.

\subsubsection{Core-dominated halo}
A halo having a volumic mass density which is constant at low radii is often 
referred to as a core dominated halo or a (pseudo-)isothermal sphere. Several formalisms 
exist to describe such a halo shape \citep[][and references therein]{bla01} 
and the following prescription is used :
\begin{equation}
\rho_{\rm iso}(R)=\frac{\rho_{0}}{\left(1 +(R/R_c)^2\right)^{1.5}},
\label{eqcore1}
\end{equation}
where $\rho_{0}$ and $R_c$ are the central mass density and the core radius of the halo. These are the parameters 
that are fitted to the rotation curve. This dark halo  was used in \citet{car06}. 
 The volumic mass density decreases at large radius like $\rho \propto R^{-3}$, exactly as for the NFW halo. 
 We refer to this halo as the ``ISO model" and pseudo-isothermal or core halo.
  
\subsection{Fittings}
\label{sec:massmodelintroduction}
 Levenberg-Marquardt least-squares fittings to the rotation curve of M31 are done.   
   One model uses \mld\ and \mlb\ fixed at the SSP values of 1.7 and 2.2 (model referred to as 
 ``SSP"). Another hybrid model (``HYB") is fitted with \mld\ $=1.7$ and \mlb\ $=0.8$ (see \S\ref{sec:stelmasstolightratio}).
 
 We have also considered best-fit models (``BF") for both photometric bands, which have free halo and stellar parameters. Only one plot is shown for this model
  given that they lead to either very unconstrained parameters (e.g. the core radius), or too massive stellar disc (the so-called ``maximum disc" 
  fitting) or unphysical models (e.g. null mass-to-light ratios). 
This later result is not acceptable for a high surface brightness galaxy like M31 and has to be rejected. 
Results of best-fit models are listed in Tab.~\ref{tab:massmodelresults} and 
 are only used to provide a very upper limit of the total stellar mass of M31.
  
A usual normal weight is applied to the data points as the inverse of the square of the velocity uncertainties. 
Several tries are done with different sets of initial guesses in order to avoid as much as possible 
 local minima. 
 Furthermore, a constraint is applied to the mass densities which have to be positive in order to avoid hollow halos. 
 Another final constraint is for the quantity $n$ of the  Einasto halo, which has to
 be larger than or equal to 0.5 because of the definition of $d_n$. 

\begin{figure*}[!t]
\includegraphics[width=0.5\textwidth]{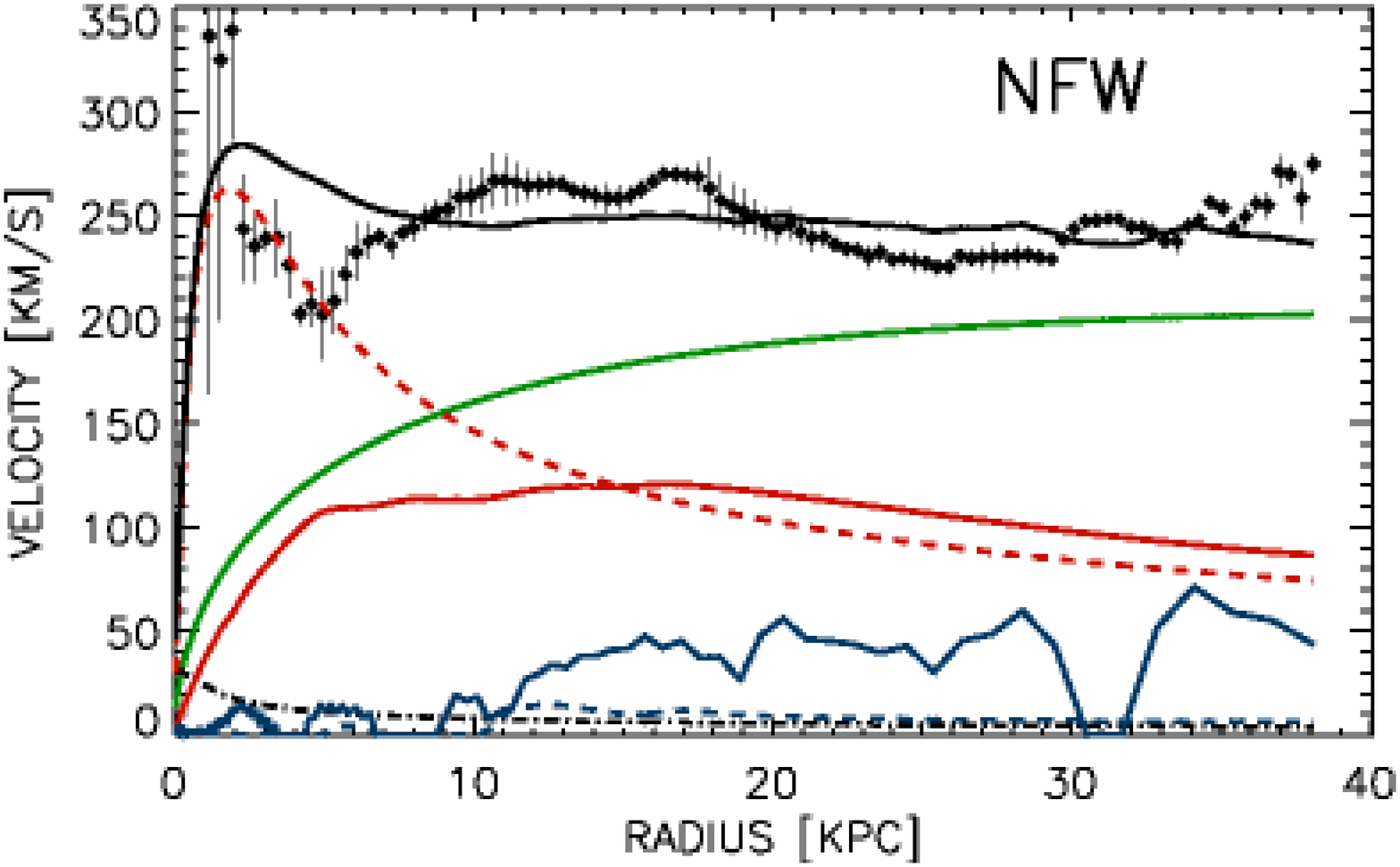}\includegraphics[width=0.5\textwidth]{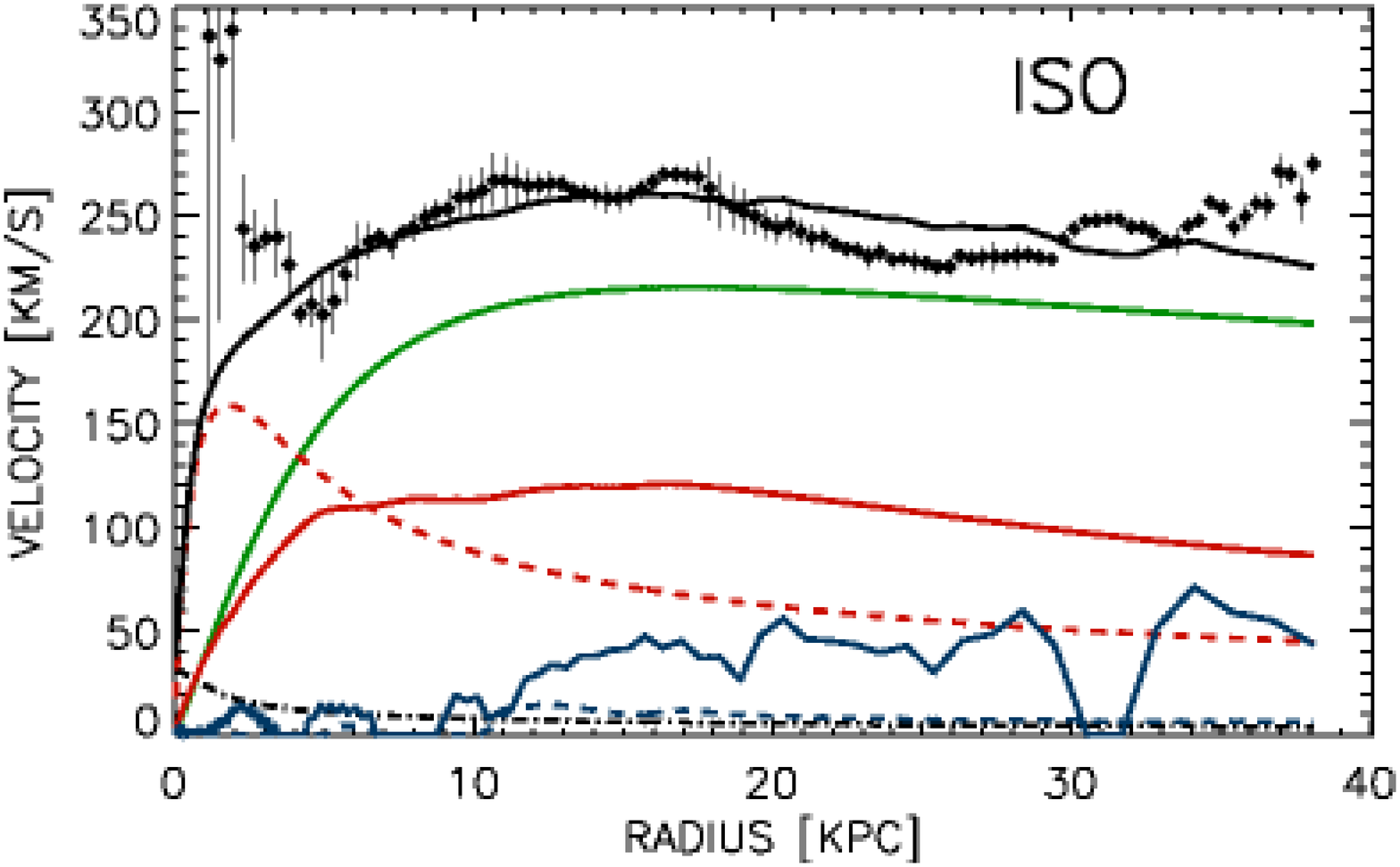}
\includegraphics[width=0.5\textwidth]{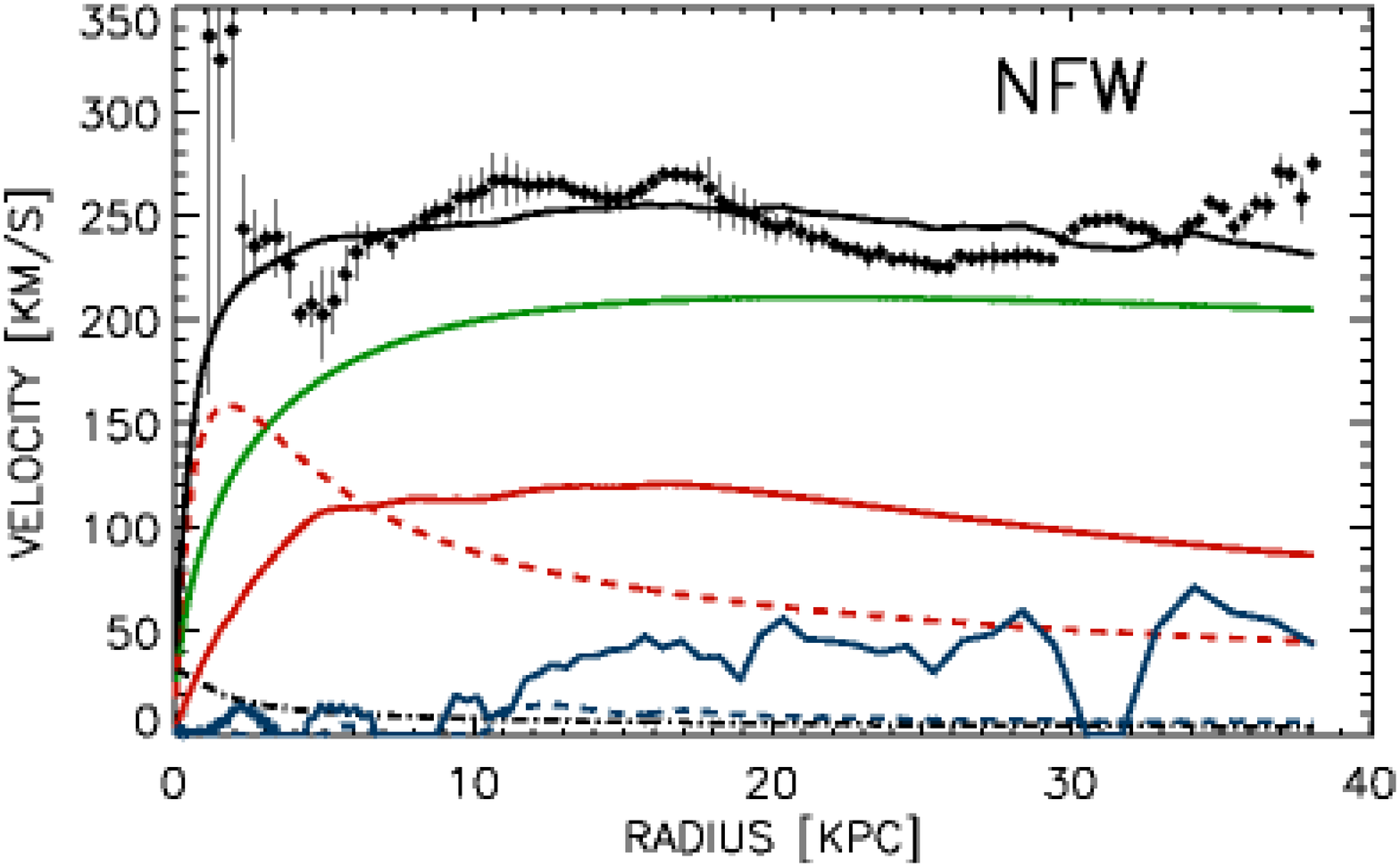}\includegraphics[width=0.5\textwidth]{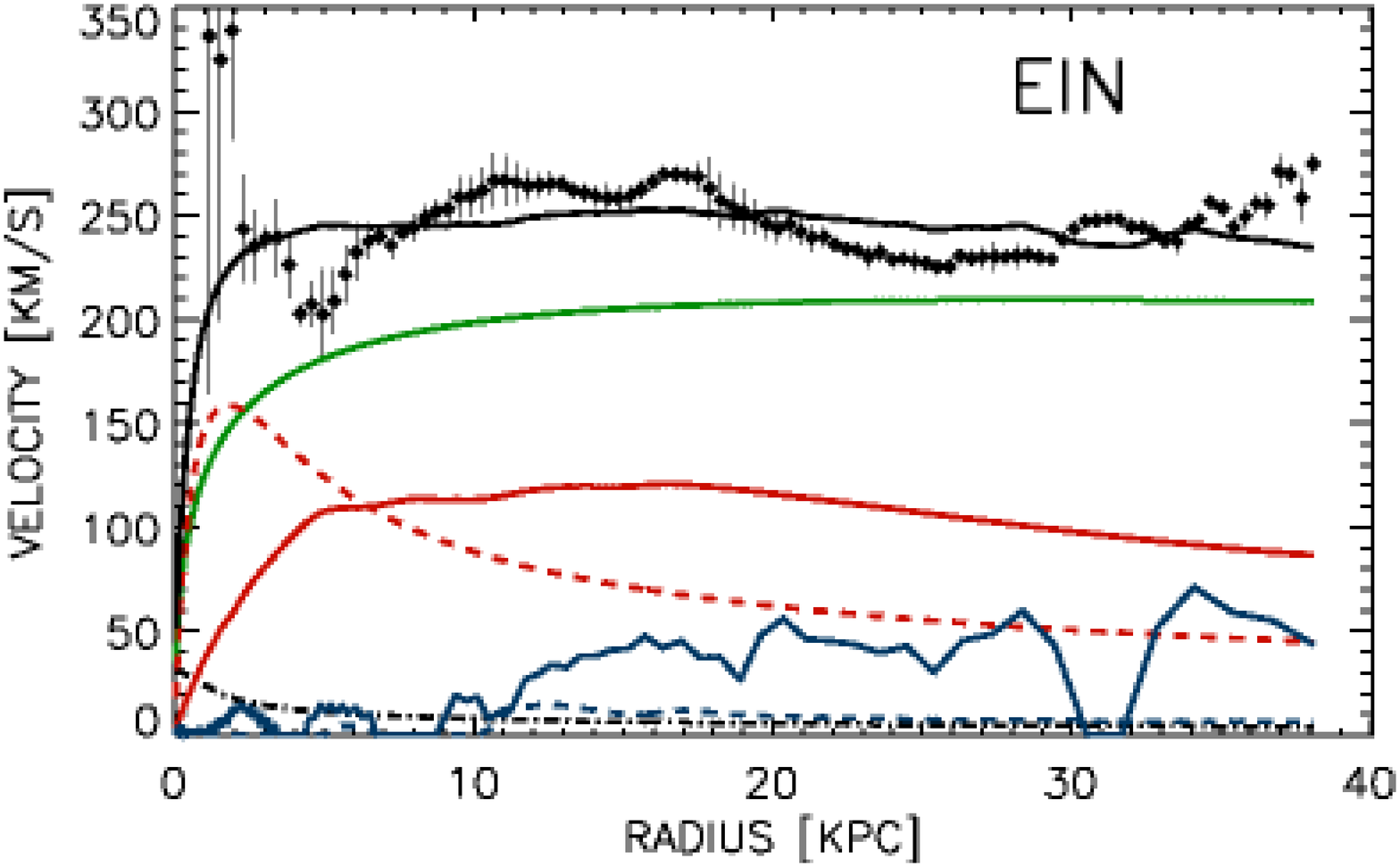}
 \caption{Mass distribution models of Messier 31.   
 \textbf{Top-Left} : Result with bulge and disc mass-to-light ratios fixed at values 
  \mlb\ $=$ 2.2 and \mld\ $=$ 1.7 (respectively) for a cuspy dark matter halo (``NFW"). \textbf{Top-right} : Same as in top-left panel
 but with   \mlb\ $=$ 0.8 for a pseudo-isothermal sphere (``ISO"). 
 \textbf{Bottom} : Same as in top-right panel but for the cusp (left) and Einasto (``EIN", right) halos.  
 A black dashed-dotted line is for the black hole contribution, solid and dashed blue lines for the neutral and molecular 
gaseous discs, solid and dashed red lines for the stellar disc and bulge,  a green line for the dark matter halo 
and a  solid black line for the overall model.}
\label{fig:bestfit1} 
\end{figure*}

\subsection{Results}
Rotation curve decompositions are displayed in Figure~\ref{fig:bestfit1} and fitted parameters listed in Tab.~\ref{tab:massmodelresults}.
 
The basic results can be summarized as:

\begin{itemize}
\item 
The reduced $\chi^2_r$ are high, it is impossible to reproduce the exact perturbed shape of the rotation curve. 
 
 \item The rotation velocities in the nucleus are not reproduced. It is expected since this region has a very negligible 
  weight in all fittings because of the large observed uncertainties.
 
  \item All models with a fixed \mlb\ $= 2.2$ 
 significantly overestimate the dip at 4 kpc, sometimes with a velocity difference of up to 60 \kms, 
 as seen in the top-left panel of Fig.~\ref{fig:bestfit1}. 

\item Hybrid models with a fixed \mlb\ $= 0.8$ gives better results than pure SSP models. 

\item  
 For each of the HYB and SSP model, the quality of the fit is equivalent whatever the shape of the dark matter halos is.

\end{itemize}

\subsection{Analysis}

High $\chi^2_r$ values are in majority caused by the peculiar shape 
of the high resolution rotation curve, which is hard to reproduce  
irrespective of the halos and mass-to-light ratios used during the fittings. From a statistical point-of-view, the derived $\chi^2_r$ 
combined with about 95 degrees of freedom indicate that the difference between the rotation curve and each dynamical model
is extremely significant, so that it could be tempting to reject any mass distribution models. 
However the purpose of this discussion is not to claim 
that the ultimate dark matter halo of M31 has been found, but rather to put basic constraints on its 
parameters and to provide a summary of the dynamical content of the galaxy. 
 
\subsubsection{Mass-to-light ratios}
\label{sec:analysismasstolight}
 We first discuss the results from SSP and HYB models.  
The worst results from all purely SSP models  are evidence that a color-based mass-to-light of 2.2 for the bulge 
has to be ruled out by the present observations unless the measured rotation curve is completely uncorrect within $R =5$ kpc ,which 
is hard to imagine. The bulge is indeed too massive in the SSP model, as seen in the top-left graph of Fig.~\ref{fig:bestfit1}. 
 
The bulge in hybrid models still overestimates the 4 kpc dip, but the difference is not as significant as for the 
SSP models. Are stellar synthesis models compatible with \mlb\ $= 0.8$?
Reconciling them  with \mlb\ $= 0.8$ would require a dust-free 
color $B-R \sim 0.62$ mag and an intrinsic reddening $E_{B-R} \sim 1$ mag towards the direction of the M31 bulge.  
However this value is 0.65 mag larger than the one adopted here. 
The global reddening law for M31 has been obtained by averaging data 
from many line-of-sights and having different dust content \citep[see eg Fig.8 of][]{bar00}. 
In view of the upper limits allowed by the uncertainties on $E_{B-R}$ and $E_{B-V}$ 
(as quoted in Barmby et al. 2000 and Williams \& Hodge 2001) 
and of  the spatial distribution of the (largest) observed reddening factors 
($E_{B-V} > 0.4$ mag),  which are not really concentrated in the bulge of M31 
but rather in the disc dust lanes, one can hardly argue that large reddenings ($E_{B-R} > 0.7$ mag) are likely 
in the direction of the bulge of M31. The adopted extinction law and galaxy color thus seem realistic.
Even by taking into account a lower value \mlb\ $\sim 1.6$, as allowed by the 
 uncertainties on $B-R$ and on the population synthesis models,  
 the fit would remain of bad quality ($\chi_r^2 = 22.4$ with the pseudo-isothermal halo for instance).
Notice that \mlb\ $\sim 1.6$ still remains at odds with the dynamical value inferred from velocity dispersion measurements. 
We are thus left at first sight to question about the validity of the 
 stellar population synthesis models of \citet{bel03} for the bulge of M31. 
We do not know the implications of a fine tuning of ingredients of 
stellar synthesis models on the stellar evolution of a bulge with \mlb\ $\sim 0.8$. 
A further analysis should be conducted to study this problem. 

Adopting \mlb\ $= 0.8$ as the most likely value implies a bulge mass-to-light ratio $\sim 2.1$ lower than the disc one. 
The stellar disc of M31 is massive but sub-maximum. At a radius $R = 2.15 R_d \sim 12$ kpc, 
which is the location of the peak velocity of an exponential disc, 
the ratio of the stellar disc velocity to the rotation curve  is $\sim 45\%$ for \mld\ $= 1.7$. 
Notice that the ratio of the total bulge$+$disc velocity to the rotation curve  rises to
$\sim 54\%$, which still shows that the stellar component is sub-maximum. 
Figure~\ref{fig:bestfit1}  shows that the dark matter component dominates everywhere 
the mass of the stellar disc, and more importantly for $\Lambda$CDM halos.

One would need \mld\ $\sim 4.8$ for the galaxy to be exactly 75\% of the maximum disc. This is the reason 
why the best-fit models correspond  to the maximum disc hypothesis because they provide the largest mass-to-light ratios. 
 There is no real concensus on the nature of discs 
to be maximum or sub-maximum since all trends are found in the observations 
and/or the numerical simulations \citep[e.g.][]{bot97,cou99,wei01,kas06,zan08}. 
A point to notice here is that the population synthesis result seems not to correspond to the 
maximum disc solution, even though in general both assumptions are thought to be tightly linked \citep{bel01}. 
The upper limit of the total stellar mass implied by the BF models is 
$2.75 \times 10^{11}$ \msol. 
Those masses strongly differ from other stellar mass estimates of M31. 
Notice that BF models give either unconstrained core radii or no stellar components, as seen in  
Tabs.~\ref{tab:massmodelresults}. 
The concentration and circular velocity of the NFW halo are also not consistent with expectations 
of similar M31-sized halos from CDM simulations  \citep{bul01, net07}. 
For all these reasons, the results of BF models are rejected in the following of the analysis.

A total stellar mass of $\M_{\rm \star} = (9.5 \pm 1.7) \times 10^{10}$ \msol\ is derived using 
\mld\ $= 1.7$ and \mlb\ $= 0.8$ and a conservative 25\% uncertainty in the disc mass-to-light ratio.  
The comparison with results of mass distribution modelings of M31 from other works \citep{ken89,car06,wid03,gee06,tem07} 
is limited because uncorrected colors or other different photometric bands have been used for 
deriving the stellar mass-to-light ratios as well as different rotation curves. 
An interesting result raised by \citet{wid03} is that most of their numerical models point out to a higher disc mass-to-light 
ratio than the bulge one, which is consistent with our current analysis.
 However \citet{gee06} deduce a higher bulge mass-to-light ratio than for the disc (3.9 and 3.3 respectively), as well as 
 \citet{tem07}. 
The total stellar mass compares very well with those inferred by \citet{wid03}, \citet{gee06} or \citet{bar07},
  but is higher than in \citet[][$\sim 7.5 \times 10^{10}$ \msol]{tem07}. \citet{wid03} 
arrive at the conclusion that the disc mass is $< 8 \times 10^{10}$ \msol\ and the bulge mass is $2.4 \times 10^{10}$ \msol, 
\citet{gee06} report $7.2  \times 10^{10}$ \msol\ for the disc and $3.2  \times 10^{10}$ \msol\ for the bulge. The disc mass we deduce 
agrees with their results within the uncertainties while the bulge only agrees well with \citet{wid03}. 
It is worth notifying that the stellar mass quoted in \citet[][$2.3 \times 10^{11}$ \msol]{car06} 
is too high compared with all other estimates. 
This result is likely attributed to the too large contribution of the bulge  and to the use 
of the best-fit/maximum disc hypothesis in our previous mass distribution model.

\subsubsection{The dark halo of M31}
\label{sec:haloshape}
The only valid hybrid models are used to discuss the basic parameters of the dark halo. 
We do not aim at determining an exact density shape of the dark halo of the Andromeda galaxy (core versus cusp). 
We only notice that CDM models describe slightly better the mass distribution than the ISO model. However this 
result is not really significant due to the high inferred $\chi^2$. 
The parameters of the Einasto halo are in good agreement 
with those of empirical galaxy-sized halos modelized in numerical simulations \citep{mer06}. 
The effective radius  $R_e$ is large and not well constrained. Actually, the Einasto model does 
not help to improve the quality of the fit, though it has one more free parameter than ISO or NFW models.

The concentration parameter of the NFW model is $c = 20.1 \pm 2.0$. This value is 
consistent with the one expected for M31-sized halos in numerical simulations 
  \citep[$8 < c < 20$][]{bul01, net07}. 
  It is 2.5 times lower than the one derived by \citet{sei08}, who used a
   rotation curve for the dynamical analysis as well.
 
\subsubsection{Dynamical content summary} 
The dynamical content of M31 is  calculated by averaging results from the hybrid model of the NFW halo. 
The total dynamical mass enclosed in the inner 38.1 kpc of M31 at the last measured data point of the \hi\ rotation curve 
 is $\M_{\rm Dyn} = (4.7 \pm 0.5) \times 10^{11}$ \msol. It represents the sum of the dark component and baryonic masses.
  It implies a  dark-to-baryonic mass ratio of
 $\M_{\rm Dark}/\M_{\rm Baryon} \sim 3.7 $ (79\% of dark matter), and a total dynamical mass-to-light ratio $\M_{\rm Dyn}/{\rm L_{tot}} \sim 6.7$. 
  Here $\M_{\rm Baryon}$ represents the sum of the black hole, gaseous and stellar masses, 
$\M_{\rm Dark}$ the dark matter mass and ${\rm L_{tot}}$ the total stellar luminosity in the $R-$band, all values integrated within $R =
38.1$ kpc. The atomic and molecular gas to stellar mass fraction is $\M_{\rm HI+He+H_2}/\M_{\rm \star} = (6.2^{+1.3}_{-1.0}) \%$.
At the virial radius of the NFW halo ($R = 159$ kpc), the total enclosed mass of M31 is 
$\M_{\rm Vir} \sim 1.0 \times 10^{12}$ \msol. 

Many works reported estimates of the total mass of M31 using different observational 
techniques to probe its potential. 
 For instance a value of $2.8 \times 10^{11}$ \msol\ inside 31 kpc has been derived by \citet{eva00a} 
 from planetary nebulae radial velocities. 
  Recent single-dish measurements of the rotation curve out to $\sim 35$ kpc led \citet{car06} to 
 the conclusion that a total mass of $3.4 \times 10^{11}$ \msol\ was enclosed inside this radius.
 These values are somewhat smaller than the new mass derived inside 38.1 kpc.
As for its total mass \citet{eva00a} have used radial velocities of fifteen dwarf satellites of M31 to estimate a value of 
$7.0^{+10.5}_{-3.5} \times 10^{11}$  \msol. 
\citet{eva00b} indicate a mass of $12.3^{+18}_{-6} \times 10^{11}$ \msol\ 
from planetary nebulae, globular clusters and galaxy satellites radial velocities measurements inside 
a very large distance to the M31 nucleus (550 kpc). 
More recent studies of the kinematics of several substructures located in the giant stellar halo 
surrounding M31 help  getting more information about the total mass of the galaxy. Among others, 
\citet{iba04} measure  the kinematics of stars inside the Giant Stream and model 
 their orbits  within a universal NFW profile. They deduce
  a total mass of $7.5^{+2.5}_{-1.3} \times 10^{11}$ inside 125 kpc, for a lower limit of $5.4 \times 10^{11}$ \msol.
With more kinematical data of RGB stars from the stellar halo, the same team indicates a higher virial mass 
$> 9 \times 10^{11}$ \msol\ \citep{cha06}. In the same time, numerical simulations of the merger with the Giant Stream progenitor
indicate a best-fit mass of $(7.4 \pm 1.2) \times 10^{11}$ \msol\ within 125 kpc \citep{far06}, very close to the result of \citet{iba04}.
Analyses of compilations of a lot of kinematical data (rotation curves and/or velocity dispersion profiles of the M31 disc) allowed \citet{gee06}
 to constrain a mass of $7.1 \times 10^{11}$ \msol, 
 $10.7 \times 10^{11}$ \msol\  inside 400 kpc for \citet{tem07}. 
 Finally, \citet{sei08} find a  virial mass of $(7.3 \pm 0.2) \times 10^{11}$ \msol\ for their NFW model 
 of another composite rotation curve of M31.
Our new extrapolated virial mass is thus in good agreement with the upper limits provided by \citet{iba04} and \citet{cha06}.

\section{Conclusion}
\label{conclusion}

A new deep \hi\ mapping of Messier 31 obtained at DRAO has been presented. It combines 
single dish antenna and high resolution synthesis telescope observations. The high spectral resolution of the 
observations show evidence that the most important contamination by the Milky Way \hi\ is done by a spectral component at
 $\sim -40$ \kms. The main results from the analysis of the neutral gas distribution, kinematics and dynamics are :

\begin{itemize}

\item Up to five \hi\ peaks are detected in spectra of the M31 disc, 
though the detection of four or five peaks remains very rare. The majority of M31 spectra exhibit 
a maximum of two \hi\ lines. The origin of all these additional lines may be internal (projection effects  
of gas in the warped part of the disc and in inner unresolved spiral arms, 
expanding clouds in star forming regions) and/or external (extraplanar gas).

\item Ring-like and spiral-like structures are observed in the gas distribution. 
This is consistent with previous \hi\ images of M31.

\item New \hi\ structures are revealed in the outskirts of the disc. 
First, thin extensions (called the \hi\ spurs)
are seen to the North-East and South-West. Their size is $\sim 7-10$ kpc. 
Evidence for velocity gradients in those structures are presented. 
 Then a faint, extended (32 kpc) and outer \hi\ structure (called the external arm) is discovered. 
Contrary to every \hi\ structures identified within the field-of-view, it has no 
morphological nor kinematical counterpart in the opposite side of the galaxy. 
Its peculiar kinematics can be explained by a $\sim 10\degr$ 
larger inclination and position angle than those of the disc.
 One of the \hi\ spur appears as a genuine kinematical extension of that external arm.

\item   A tight relationship is evidenced between the \hi\ spurs and perturbed stellar structures (the 
 ``G1" and north-eastern clumps). 
 As possible radial outflowing motions are detected in the \hi\ spurs, we argue that 
the perturbed stellar and gaseous outskirts are being torn from the M31 disc. 
Furthermore the external arm could be the vestige of a past tidal interaction, or even perhaps gas accretion onto the M31 disc.

\item The rotation curve is peculiar. In particular it shows a central velocity dip at 4 kpc and a rising shape 
in its outer part. A peak up to 340 \kms\ is observed within the inner 2 kpc. 
Except for the inner regions, the axisymmetry of the gas rotation  between 
the two disc halves is very good. The rotation velocity uncertainties are mostly below 10 \kms. 

\item Between $R= 6$ kpc and $R=27$ kpc a mean 
inclination  of $(74.3 \pm 1.1)\degr$ is derived for the \hi\ disc of M31, for a mean position angle of $(37.7 \pm 0.9)\degr$.
A prominent \hi\ warp is detected inside  $R = 6$ kpc, where the disc appears less inclined, while  
another external warp detected beyond $R = 27$ kpc makes the disc being less inclined and then more inclined. 
The disc tilting is accompanied by a twist of the kinematical major axis.  
 
\item  The peculiar shape of the rotation curve is hardly reproduced by mass distribution models. 
The central velocity dip cannot be modelled by any fittings. The bulge mass-to-light ratio  
deduced from color-based stellar synthesis models appropriate. 
A value close to the expected one from stellar velocity dispersion measurements provides better results.    

\item A total stellar mass of $\M_{\rm \star} = (9.5 \pm 1.7) \times 10^{10}$ \msol\ is found. 
A dynamical enclosed mass of $\M_{\rm Dyn} = (4.7 \pm 0.5) \times 10^{11}$ \msol\ is derived 
at the last measured radius of the rotation curve ($R = 38$ kpc). 
A dark-to-baryonic mass ratio of $\sim 4$ and a total gas-to-stellar mass fraction of $(6.2^{+1.3}_{-1.0}) \%$ are also derived 
inside that radius.
The total mass of M31 extrapolated to its virial radius ($R = 159$ kpc) is $\M_{\rm Vir} \sim 1.0 \times 10^{12}$ \msol.
Both these values are consistent with those given by other works using other dynamical tracers or methods. 
 
\end{itemize}

Several future works should be conducted with these data.
 A first crucial point will be to investigate the origin and evolution of the perturbed \hi\ outskirts - the external arm and spurs.  
 With the help of future N-body and hydrodynamical 
 simulations, the \hi\ kinematics should put new constraints on the mass assembly of M31. 
A second challenge will be to understand the origin of the peculiar velocities in the central 4 kpc. The perturbed velocities 
are related to the presence of the inner warp, which has perhaps been generated 
by an interaction with the small companion M32 (F. Combes \& F. Bournaud, private communication) 
and/or by the bar/boxy bulge perturbation. 
Another work will focus on a 3D modeling of the datacube in order 
to study the origin of the gas components which are seen in addition to the warped disc of M31. 
Whether M31 contains the  ``anomalous" gas from a lagging halo, as it is observed 
in recent deep \hi\ observations of other galaxies, is a question that should be addressed 
in such a model. 
 
These new results obtained for M31 could only be obtained 
by a combination of deep exposures with high spectral resolution measurements. The quest of understanding 
the formation and evolution of M31 should benefit from such deep observations.
We expect that future radio observatories like 
 the Square Kilometre Array (in 2016) and its australian and south-african precursors (in 2013) will renew 
 the vision we have of the outer \hi\ distribution and dynamics for millions of galaxies in the local Universe. 
 For the first time they will also allow to trace the evolution of the atomic gas as a function of redshift 
  and thus the growth of the baryonic mass of Milky Way- and M31-like galaxies. 

\section*{Acknowledgements}
 We are very grateful to the DRAO staff for their support in
obtaining those observations, and especially T. Landecker for
encouraging us to pursue this project. 
We are grateful to an anonymous referee whose  critical
comments helped improving the data analysis and the clarity of the article.
We are grateful to R. Braun, G. Mamon, as well as M. Lehnert and W. van Driel at G\'EPI  
 for fruitful discussions, to R. Kothes for thoughtful comments on the
manuscript, to P. Barmby for providing us with the \textit{Spitzer/IRAC} photometric
profile of M31 and to C. Willmer for his help in
deriving the absolute magnitude of the Sun for the \textit{Spitzer/IRAC} camera ($\rm M_{Sun}^{3.6} = 3.292$).
 L. C. and C. C. acknowledge financial support from CRSNG, Canada and FQRNT,
Qu\'ebec  and from the European Community Framework Programme 6 - Square Kilometre Array Design
Studies (SKADS).  This work has partially been supported by a grant from the
Brandon University Research Committee (BURC) to T. F.

\appendix
\begin{deluxetable}{c||c||c|c|c}
 \tablecaption{Mass model results using the $R-$band photometry.}
\tablehead{ Halo model &  Parameter   & \colhead{SSP} & \colhead{HYB} & \colhead{BF} }
\startdata
ISO & $\rho_0$ & 28.9 $\pm$   4.9 & 83.1  $\pm$  13.3  & 1.2 $\pm$ 0.1 \\
~ & $r_c$ &  9.3  $\pm$  0.8  & 6.0  $\pm$  0.4   & 1347.3   $\pm$ 104481  \\
~ &  \mld\ & 1.7  & 1.7     & 5.9 $\pm$   0.3 \\
~ &  \mlb\ & 2.2  & 0.8     & 1.0  $\pm$  0.2 \\
~ &  $\chi^2_r$ & 23.6  &  22.9     &  16.6    \\
~ &  ~ & ~& ~ \\
NFW & $V_{200}$ & 168.5 $\pm$  9.2  &  146.2   $\pm$  3.9  &  462.7    $\pm$ 52.6  \\
~ & $c$ & 10.1  $\pm$   1.4  &  20.1   $\pm$   2.0  &  0.9    $\pm$   0.1\\
~ & \mld\ &   1.7  & 1.7    &	5.1   $\pm$  0.3  \\
~ & \mlb\ &   2.2  & 0.8    &	1.1   $\pm$    0.3  \\
~ &  $\chi^2_r$ & 23.6  & 21.0  &  19.7 \\
~ &  ~ & ~& ~ \\
EIN & $\rho_e$ & 0.048  $\pm$ 0.196 & 0.001 $\pm$   0.007   & 0.508 $\pm$   1.083 \\
~ &  $R_e$ & 99.3 $\pm$ 174.5 &  426.0 $\pm$ 1222.7 & 34.9 $\pm$ 27.5  \\
~ &  $n$ & 3.5   $\pm$   2.7   &  7.9   $\pm$  5.2  &  4.0    $\pm$    1.1  \\
~ & \mld\ &   1.7  & 1.7    &	0.001 $\pm$ 0.080 \\
~ & \mlb\ &   2.2  & 0.8    &	0.013 $\pm$ 0.857 \\
~ &  $\chi^2_r$ &  23.8  & 20.7  & 21.2   \\
\enddata
\tablecomments{Radii are in kpc, volumic mass densities in $10^{-3}$ \msol\ pc$^{-3}$, velocities in \kms\ and
mass-to-light ratios in \msol/\lsol. The 
reduced chi-square $\chi^2_r$ is given for each model.}
\label{tab:massmodelresults}
\end{deluxetable}


\begin{deluxetable}{ccccccccccc}
\tablecaption{\hi\ rotation curve, surface density, 
 disc inclination and position angle from tilted-ring model results.}
\tablehead{ \colhead{Radius} &  \colhead{Radius} &  \colhead{\pa} &  \colhead{$e_{P.A.}$}  & 
\colhead{\pa$_{\rm adopted}$} &\colhead{$i$} &  \colhead{$e_{i}$}  & 
\colhead{$i_{\rm adopted}$} & \colhead{\vrot} & \colhead{$e_{v_{\rm rot}}$} & \colhead{$\rm \Sigma_{HI}$} \\
\colhead{(\arcmin)} &  \colhead{(kpc)} & \colhead{(\degr)} & \colhead{(\degr)} & 
\colhead{(\degr)} & \colhead{(\degr)} & \colhead{(\degr)} & \colhead{(\degr)}  & 
\colhead{(\kms)} & \colhead{(\kms)} &   \colhead{(\msol\ pc$^{-2}$)}}
\startdata
  1.67 &    0.38  & \nodata &\nodata	      &     26.6   & \nodata	       & \nodata	  &    31.0  & \nodata        & \nodata	   & 	2.77       \\	  
  3.33 &    0.76  & \nodata	   & \nodata	      &     26.6   & \nodata	       & \nodata	  &    31.0  &  \nodata       & \nodata	   & 	3.19       \\	  
  5.00 &    1.14  & \nodata	   & \nodata	      &     26.6   & \nodata	       & \nodata	  &    31.0  &  336.2  &    171.7  & 	4.21      \\	  
  6.67 &    1.52  & \nodata	   & \nodata	      &     26.6   & \nodata	       & \nodata	  &    31.0  &  324.6  &    125.1  & 	3.94      \\	  
  8.33 &    1.90  &  26.6  & 	 2.5  &     26.8   & 	 30.5  &     4.3  &    31.0  &  339.0  &     52.8  & 	4.49      \\	  
 10.00 &    2.28  &  27.1  & 	 2.2  &     27.3   & 	 62.4  &     4.3  &    51.0  &  243.6  &     25.8  & 	2.14      \\	  
 11.67 &    2.66  &  27.7  & 	 1.9  &     28.6   & 	 61.0  &     3.4  &    61.3  &  235.2  &     17.0  & 	1.52      \\	  
 13.33 &    3.04  &  30.7  & 	 1.5  &     30.6   & 	 66.3  &     2.7  &    65.8  &  238.9  &      5.7  & 	1.51      \\	  
 15.00 &    3.43  &  33.5  & 	 1.3  &     32.2   & 	 70.4  &     3.0  &    68.1  &  239.3  &     18.3  & 	1.57     \\	  
 16.67 &    3.81  &  32.6  & 	 1.2  &     32.7   & 	\nodata       & \nodata	  &    67.7  &  226.3  &     16.1  & 	2.21     \\	  
 18.33 &    4.19  &  32.3  & 	 1.0  &     32.9   & 	 69.3  &     1.5  &    67.3  &  202.6  &      4.7  & 	3.00     \\	  
 20.00 &    4.57  & \nodata	   & \nodata	      &     33.3   & 	 62.5  &     1.0  &    64.4  &  207.3  &     10.7  & 	3.54      \\	  
 21.67 &    4.95  & \nodata	   & \nodata	      &     33.6   & 	 60.8  &     1.2  &    63.7  &  202.5  &     21.7  & 	2.75      \\	  
 23.33 &    5.33  &  33.4  & 	 0.9  &     33.9   & 	 66.8  &     1.3  &    65.9  &  208.9  &     15.6  & 	2.31      \\	  
 25.00 &    5.71  &  36.0  & 	 0.5  &     35.4   & 	 69.7  &     1.4  &    68.1  &  221.6  &     13.4  & 	2.09      \\	  
 26.67 &    6.09  &  37.0  & 	 0.6  &     36.4   & 	 66.8  &     0.7  &    69.7  &  232.2  &     13.7  & 	1.86      \\	  
 28.33 &    6.47  &  36.3  & 	 0.5  &     36.6   & 	 74.3  &     0.8  &    72.0  &  237.6  &      8.3  & 	1.41      \\	  
 30.00 &    6.85  &  36.9  & 	 0.4  &     36.5   & 	 73.2  &     0.4  &    73.5  &  239.8  &      2.2  & 	1.43      \\	  
 31.67 &    7.23  &  35.7  & 	 0.3  &     36.4   & 	 74.6  &     0.6  &    74.3  &  235.6  &      6.1  & 	1.60      \\	  
 33.33 &    7.61  &  37.0  & 	 0.2  &     36.8   & 	 75.0  &     0.4  &    74.6  &  241.7  &      3.3  & 	2.07      \\	  
 35.00 &    7.99  &  37.2  & 	 0.2  &     37.3   & 	 74.6  &     0.4  &    74.5  &  244.3  &      6.4  & 	2.84      \\	  
 36.67 &    8.37  &  37.9  & 	 0.2  &     37.7   & 	 73.7  &     0.4  &    74.3  &  248.8  &      6.4  & 	3.85      \\	  
 38.33 &    8.75  &  38.1  & 	 0.3  &     38.0   & 	 74.8  &     0.4  &    74.3  &  251.8  &      5.5  & 	4.61     \\	  
 40.00 &    9.13  &  38.0  & 	 0.3  &     38.3   & 	 73.8  &     0.5  &    74.4  &  253.0  &      9.2  & 	4.39      \\	  
 41.67 &    9.51  &  38.9  & 	 0.2  &     38.7   & 	 75.0  &     0.5  &    74.8  &  258.8  &      9.6  & 	3.78      \\	  
 43.33 &    9.90  &  39.1  & 	 0.2  &     39.0   & 	 75.2  &     0.4  &    75.2  &  259.0  &      9.5  & 	3.34      \\	  
 45.00 &   10.28  &  39.2  & 	 0.2  &     39.1   & 	 75.3  &     0.4  &    75.6  &  262.2  &     10.8  & 	3.16      \\	  
 46.67 &   10.66  &  39.1  & 	 0.2  &     39.0   & 	 76.7  &     0.4  &    76.1  &  266.8  &     13.0  & 	3.37      \\	  
 48.33 &   11.04  &  38.7  & 	 0.2  &     38.8   & 	 76.1  &     0.3  &    76.3  &  266.8  &     11.7  & 	4.34       \\	  
 50.00 &   11.42  &  38.5  & 	 0.2  &     38.6   & 	 76.6  &     0.3  &    76.4  &  265.9  &      9.9  & 	5.02       \\	  
 51.67 &   11.80  &  38.7  & 	 0.2  &     38.3   & 	 76.5  &     0.3  &    76.3  &  264.4  &      7.2  & 	4.95       \\	  
 53.33 &   12.18  &  37.8  & 	 0.2  &     37.7   & 	 76.1  &     0.3  &    76.0  &  264.7  &      5.3  & 	4.45       \\	  
 55.00 &   12.56  &  36.8  & 	 0.2  &     37.0   & 	 75.5  &     0.3  &    75.6  &  265.3  &      5.4  & 	4.03      \\	  
 56.67 &   12.94  &  36.1  & 	 0.2  &     36.4   & 	 75.1  &     0.3  &    75.1  &  265.2  &      3.9  & 	3.98       \\	  
 58.33 &   13.32  &  36.3  & 	 0.2  &     36.1   & 	 74.7  &     0.3  &    74.8  &  262.0  &      2.4  & 	3.91	    \\     
 60.00 &   13.70  &  35.5  & 	 0.2  &     35.9   & 	 74.7  &     0.3  &    74.5  &  260.8  &      4.5  & 	3.67	    \\     
 61.67 &   14.08  &  36.0  & 	 0.2  &     36.1   & 	 73.9  &     0.4  &    74.3  &  259.2  &      5.4  & 	3.39	   \\	   
 63.33 &   14.46  &  36.6  & 	 0.2  &     36.6   & 	 74.9  &     0.3  &    74.1  &  258.1  &      6.8  & 	3.37	   \\	   
 65.00 &   14.84  &  37.3  & 	 0.2  &     37.1   & 	 72.9  &     0.3  &    73.7  &  258.4  &      6.1  & 	3.01	   \\	   
 66.67 &   15.23  &  37.3  & 	 0.2  &     37.3   & 	 73.7  &     0.3  &    73.7  &  259.2  &      4.6  & 	2.61	   \\	   
 68.33 &   15.61  &  37.4  & 	 0.2  &     37.2   & 	 73.9  &     0.3  &    74.1  &  262.7  &      4.2  & 	2.39	   \\	   
 70.00 &   15.99  &  36.8  & 	 0.2  &     37.1   & 	 74.5  &     0.3  &    74.7  &  266.1  &      4.0  & 	2.00	   \\	   
 71.67 &   16.37  &  37.2  & 	 0.2  &     37.3   & 	 75.8  &     0.3  &    75.4  &  270.0  &      2.3  & 	1.92	  \\	   
 73.33 &   16.75  &  37.9  & 	 0.2  &     37.7   & 	 75.7  &     0.3  &    75.5  &  269.8  &      1.0  & 	1.79	  \\	   
 75.00 &   17.13  &  38.1  & 	 0.2  &     37.7   & 	 75.6  &     0.3  &    75.1  &  269.1  &      3.9  & 	2.21	  \\	   
 76.67 &   17.51  &  37.1  & 	 0.2  &     37.5   & 	 73.7  &     0.4  &    74.3  &  268.5  &      7.1  & 	1.91	   \\	   
 78.33 &   17.89  &  37.5  & 	 0.2  &     37.5   & 	 74.2  &     0.5  &    73.8  &  263.0  &     15.0  & 	1.74	   \\	   
 80.00 &   18.27  &  37.7  & 	 0.3  &     37.5   & 	 72.8  &     0.5  &    73.4  &  257.1  &     13.5  & 	1.60	   \\	   
 81.67 &   18.65  &  37.1  & 	 0.3  &     37.4   & 	 73.4  &     0.5  &    73.4  &  254.1  &     10.6  & 	1.44	   \\	   
 83.33 &   19.03  & \nodata	   & \nodata	      &     37.5   & 	 73.8  &     0.5  &    73.6  &  251.9  &     11.1  & 	1.35	   \\	   
 85.00 &   19.41  &  37.1  & 	 0.3  &     37.7   & 	 73.7  &     0.5  &    73.4  &  249.5  &      8.1  & 	1.24	   \\	   
 86.67 &   19.79  &  38.8  & 	 0.3  &     38.3   & 	\nodata       & \nodata	  &    73.2  &  245.7  &      7.4  & 	1.15	   \\	   
 88.33 &   20.18  &  39.2  & 	 0.3  &     38.6   & 	 72.5  &     0.6  &    73.0  &  243.7  &      6.6  & 	1.18	   \\	   
 90.00 &   20.56  &  37.9  & 	 0.3  &     38.3   & 	\nodata       & \nodata	  &    73.0  &  245.9  &      7.5  & 	1.10	   \\	   
 91.67 &   20.94  &  37.7  & 	 0.3  &     38.2   & 	\nodata       & \nodata	  &    73.0  &  242.3  &      6.1  & 	1.08	   \\	   
 93.33 &   21.32  &  38.9  & 	 0.3  &     38.4   & 	 73.0  &     0.5  &    73.0  &  239.2  &      6.3  & 	0.83	   \\	   
 95.00 &   21.70  &  38.3  & 	 0.3  &     38.2   & 	 73.0  &     0.6  &    73.3  &  239.5  &      4.7  & 	0.75	  \\	   
 96.67 &   22.08  &  37.5  & 	 0.3  &     37.8   & 	 74.3  &     0.6  &    73.5  &  236.1  &      1.8  & 	0.76	   \\	   
 98.33 &   22.46  &  37.4  & 	 0.3  &     37.5   & 	 72.9  &     0.4  &    73.4  &  233.8  &      1.7  & 	0.88	   \\	   
100.00 &   22.84  &  37.5  & 	 0.3  &     37.6   & 	 73.3  &     0.5  &    73.3  &  233.1  &      3.3  & 	0.82	   \\	   
101.67 &   23.22  &  38.0  & 	 0.3  &     37.7   & 	 73.6  &     0.5  &    73.3  &  230.1  &      5.5  & 	0.75	   \\	   
103.33 &   23.60  &  37.4  & 	 0.3  &     37.7   & 	 72.9  &     0.5  &    73.2  &  232.1  &      5.0  & 	0.67	   \\	   
105.00 &   23.98  &  37.6  & 	 0.3  &     37.7   & 	 73.4  &     0.6  &    73.2  &  228.7  &      1.7  & 	0.58	    \\     
106.67 &   24.36  & \nodata	   & \nodata	      &     37.8   & 	 73.0  &     0.6  &    73.2  &  229.1  &      1.8  & 	0.49	    \\     
108.33 &   24.75  &  38.0  & 	 0.3  &     38.0   & 	 73.5  &     0.6  &    73.4  &  227.9  &      5.0  & 	0.43	    \\     
110.00 &   25.13  & \nodata	   & \nodata	      &     38.0   & 	 73.8  &     0.6  &    73.6  &  226.9  &      2.0  & 	0.42	    \\     
111.67 &   25.51  &  38.5  & 	 0.3  &     38.1   & 	\nodata       & \nodata	  &    73.6  &  225.1  &      1.6  & 	0.45	   \\	   
113.33 &   25.89  &  37.6  & 	 0.3  &     38.1   & 	\nodata       & \nodata	  &    73.7  &  225.4  &      1.8  & 	0.46	    \\     
115.00 &   26.27  &  38.4  & 	 0.3  &     38.3   & 	 73.1  &     0.5  &    73.7  &  230.3  &      2.0  & 	0.42	     \\     
116.67 &   26.65  & \nodata	   & \nodata	      &     38.4   & 	 74.5  &     0.5  &    74.0  &  229.0  &      2.3  & 	0.39	     \\     
118.33 &   27.03  &  38.9  & 	 0.3  &     38.5   & 	 74.2  &     0.5  &    73.9  &  229.9  &      4.8  & 	0.38	    \\      
120.00 &   27.41  &  38.0  & 	 0.3  &     38.4   & 	 73.3  &     0.7  &    73.5  &  230.1  &      6.6  & 	0.29	    \\      
121.67 &   27.79  &  38.5  & 	 0.3  &     38.2   & 	 72.4  &     0.9  &    73.6  &  229.8  &      3.0  & 	0.23	    \\      
123.33 &   28.17  &  37.8  & 	 0.3  &     37.8   & 	 75.6  &     0.5  &    74.0  &  230.4  &      5.2  & 	0.20	    \\      
125.00 &   28.56  &  37.2  & 	 0.3  &     37.4   & 	 73.7  &     0.9  &    73.5  &  230.9  &      2.9  & 	0.21	    \\      
126.67 &   28.94  &  37.0  & 	 0.3  &     37.0   & 	 71.5  &     1.1  &    72.5  &  229.8  &      2.1  & 	0.17	    \\      
128.33 &   29.32  &  36.6  & 	 0.5  &     36.4   & 	 73.3  &     1.5  &    71.3  &  228.8  &      1.8  & 	0.10	   \\	    
130.00 &   29.70  &  36.2  & 	 0.4  &     34.9   & 	 68.2  &     2.0  &    69.3  &  238.3  &      3.3  & 	0.08	   \\	    
131.67 &   30.08  &  31.7  & 	 1.3  &     32.5   & 	 67.8  &     1.8  &    67.2  &  243.6  &      1.4  & 	0.07	   \\	    
133.33 &   30.46  &  29.6  & 	 1.7  &     31.0   & 	 64.2  &     2.5  &    65.7  &  247.3  &      3.1  & 	0.07	    \\      
135.00 &   30.84  & \nodata	   & \nodata	      &     31.1   & 	 66.2  &     1.6  &    64.8  &  247.8  &      1.3  & 	0.10	    \\      
136.67 &   31.22  &  31.4  & 	 6.0  &     31.3   & 	 61.7  &     3.5  &    64.8  &  248.4  &      2.0  & 	0.08	    \\      
138.33 &   31.61  & \nodata	   & \nodata	      &     31.6   & 	 67.7  &     1.6  &    66.6  &  248.1  &      1.5  & 	0.07	    \\      
140.00 &   31.99  & \nodata	   & \nodata	      &     32.0   & 	 69.1  &     1.1  &    68.7  &  244.5  &      1.6  & 	0.06	    \\      
141.67 &   32.37  & \nodata	   & \nodata	      &     32.3   & 	 70.4  &     1.3  &    70.1  &  244.4  &      3.0  & 	0.06	    \\      
143.33 &   32.75  &  31.4  & 	 1.8  &     32.6   & 	 69.7  &     2.5  &    72.1  &  241.7  &      4.3  & 	0.06	    \\      
145.00 &   33.13  &  36.0  & 	 0.3  &     34.5   & \nodata	       & \nodata	  &    72.9  &  237.7  &      1.9  & 	0.02	    \\      
146.67 &   33.51  &  35.3  & 	 0.4  &     35.8   & \nodata	       & \nodata	  &    73.7  &  237.6  &      6.1  & 	0.01	    \\      
148.33 &   33.89  & \nodata	   & \nodata	      &     36.2   & 	       & 	  &    74.5  &  244.9  &      3.6  & 	0.01	    \\      
150.00 &   34.27  &  37.3  & 	 0.3  &     36.5   & 	 77.3  &     1.5  &    75.2  &  247.9  &      3.2  & 	0.01	    \\      
151.67 &   34.66  &  36.8  & 	 0.6  &     36.5   & 	 78.0  &     0.8  &    77.3  &  256.3  &      3.1  & 	0.02	   \\	    
153.33 &   35.04  &  35.8  & 	 0.5  &     36.3   & 	 78.1  &     1.2  &    77.6  &  253.5  &      4.1  & 	0.03	    \\      
155.00 &   35.42  & \nodata	   & \nodata	      &     36.3   & 	 76.5  &     3.0  &    77.3  &  244.3  &      4.7  & 	0.02	    \\      
156.67 &   35.80  & \nodata	   & \nodata	      &     36.3   & 	 77.2  &     1.8  &    77.7  &  249.3  &      5.8  & 	0.02	    \\      
158.33 &   36.18  & \nodata	   & \nodata	      &     36.3   & 	       & 	  &    78.2  &  255.7  &      4.5  & 	0.02	    \\      
160.00 &   36.56  &  36.2  & 	 0.4  &     36.3   & 	 78.9  &     1.1  &    78.7  &  255.0  &      5.8  & 	0.02	    \\      
161.67 &   36.94  &  36.7  & 	 0.9  &     36.5   & 	 80.0  &     0.7  &    79.4  &  271.1  &      7.8  & 	0.03	     \\     
163.33 &   37.32  &  34.8  & 	 2.0  &     36.6   & 	 79.3  &     0.9  &    79.5  &  269.8  &      4.7  & 	0.03	     \\     
165.00 &   37.71  &  36.0  & 	 0.6  &     36.7   & 	 79.4  &     1.4  &    79.4  &  258.2  &     10.7  & 	0.03	     \\     
166.67 &   38.09  &  36.6  & 	 1.4  &     36.7   & 	 79.3  &     1.2  &    79.4  &  275.1  &      4.8  & 	0.02	     \\       
\enddata
 \label{tab:rotcur}
\tablenotetext{a}{Innermost inclination from \citet{bra91}}
 \end{deluxetable}

\end{document}